%% file: main.tex
\documentclass[letterpaper,twocolumn,10pt]{article}
\usepackage{usenix-2020-09}
\usepackage[cmex10]{amsmath}
\usepackage{algorithmic}
\usepackage[numbers,sort]{natbib}
\usepackage[pdftex]{graphicx}
\usepackage[utf8]{inputenc}
\usepackage{enumitem}
\usepackage{multirow}
\usepackage{textcomp}
\usepackage{xcolor}
\usepackage{booktabs}
\usepackage{pifont}
\usepackage{subcaption}
\usepackage{amssymb,amsfonts}
\usepackage{balance}
\usepackage{wasysym}

\usepackage{authblk} 

\let\oldbibitem\bibitem
\def\bibitem{\vfill\oldbibitem}
\providecommand{\myparab}[1]{\smallskip\noindent\textbf{#1}\hspace{0.6em}}
\newcommand{\sectionref}[1]{$\S$\ref{#1}}
\providecommand{\sysname}{\textit{\texttt{GFWatch}}}

\hypersetup{
	colorlinks	= true,
	citecolor	= black,
	filecolor	= black,
	linkcolor	= black,
	urlcolor	= black,
	pagebackref	= false,
}

\begin{document}
\date{}

\title{How Great is the Great Firewall? Measuring China's DNS Censorship}

\author{
	Nguyen Phong Hoang$^{\star\dagger}$ \hspace{.5em}
	Arian Akhavan Niaki$^{\mathsection}$ \hspace{.5em}
	Jakub Dalek$^{\dagger}$ \hspace{.5em}
	Jeffrey Knockel$^{\dagger}$ \hspace{.5em}\\
	Pellaeon Lin$^{\dagger}$ \hspace{.5em}
	Bill Marczak$^{\dagger\mathparagraph}$ \hspace{.5em}
	Masashi Crete-Nishihata$^{\dagger}$ \hspace{.5em}
	Phillipa Gill$^{\mathsection}$ \hspace{.5em}
	Michalis Polychronakis$^{\star}$
	}
\affil{
	$^{\star}$Stony Brook University, New York, USA \hspace{1em} $^\mathsection$University of Massachusetts, Amherst, USA\\
	$^\dagger$Citizen Lab, University of Toronto, Canada \hspace{1em} $^\mathparagraph$University of California, Berkeley, USA
}

\maketitle

\input{abstract}
\input{intro}

\input{background}

\input{method}
\input{censored_domains}
\input{forged_ips}
\input{censorship_leakage}
\input{circumvention}
\input{discussion}
\input{related}
\input{conclusion}
\input{ack}

{\footnotesize
\bibliographystyle{plain}
\balance
\bibliography{main}
}
\input{appendix}

\end{document}

%% file: abstract.tex
\begin{abstract}

The DNS filtering apparatus of China's Great Firewall (GFW) has evolved
considerably over the past two decades. However, most prior studies of China's
DNS filtering were performed over short time periods, leading to unnoticed
changes in the GFW's behavior. In this study, we introduce \sysname, a
large-scale, longitudinal measurement platform capable of testing hundreds of
millions of domains daily, enabling continuous monitoring of the GFW's DNS
filtering behavior.

We present the results of running \sysname\ over a nine-month period, during
which we tested an average of 411M domains per day and detected a total of
311K domains censored by GFW's DNS filter. To the best of our knowledge, this
is the largest number of domains tested and censored domains discovered in the
literature. We further reverse engineer regular expressions used by the GFW
and find 41K innocuous domains that match these filters, resulting in
overblocking of their content. We also observe bogus IPv6 and globally
routable IPv4 addresses injected by the GFW, including addresses owned by US
companies, such as Facebook, Dropbox, and Twitter.

Using data from \sysname, we studied the impact of GFW blocking on the global
DNS system. We found 77K censored domains with DNS resource records polluted
in popular public DNS resolvers, such as Google and Cloudflare.  Finally, we
propose strategies to detect poisoned responses that can (1) sanitize poisoned
DNS records from the cache of public DNS resolvers, and (2) assist in the
development of circumvention tools to bypass the GFW's DNS censorship.

\end{abstract}

%% file: intro.tex
\section{Introduction}
\label{sec:intro}

Among the censorship regimes on the Internet, China is one of the most
notorious, having developed an advanced filtering system, known as the Great
Firewall (GFW), to control the flow of online information. The GFW's worldwide
reputation~\cite{FreedomHouse2018} and ability to be measured from outside the
country, has drawn the attention of researchers from various disciplines,
ranging from political science~\cite{Chase2002YouveGD, Deibert2002DarkGA,
Bambauer2005InternetFI, Deibert10chinacyberspace} to information and computer
science~\cite{Winter2012, china:2014:dns:anonymous, Marczak2015AnAO,
Ensafi2015, Arun:foci18, Tripletcensors}.

Unlike many other DNS censorship approaches, the GFW is known to return
globally routable IP addresses in its injected responses. Recent
studies~\cite{hoang:2019:measuringI2P, Hoang2020:ASIACCS, Tripletcensors} have
observed injected IP addresses belonging to popular US companies, including
Facebook, Dropbox, and Twitter. The use of routable IPs is in contrast to
countries such as  Bahrain, Korea, Kuwait, Iran, Oman, Qatar, Thailand, or
Yemen~\cite{Jones2014AutomatedDA, Gebhart2017InternetCI,
hoang:2019:measuringI2P, ICLab:SP20, Raman2020MeasuringTD}, where DNS
censorship redirects users to blockpages that inform users about the blocked
content. It is also in contrast to censors using fixed DNS responses such as
NXDOMAIN~\cite{rfc8020, pakistancensorship, Iris, ICLab:SP20} or addresses
from private IP ranges (e.g., 10.0.0.0/8)~\cite{syriacensorship, Aryan:2013,
Iris}. This use of globally routable IPs by the GFW has implications for
censorship detection, which needs to carefully distinguish censored from
legitimate DNS responses, and also makes detecting and mitigating leaked DNS
responses from public resolvers non-trivial.

Despite the many previous studies that examine the technical strategies
employed by the GFW, such as TCP/IP packet
filtering~\cite{Clayton2006IgnoringTG, Park2010EmpiricalSO, Winter2012,
ensafi2015analyzing, Arun:foci18} and DNS poisoning~\cite{Sparks:CCR2012,
china:2014:dns:anonymous, holdonDNS, farnan2016cn.poisoning}, there has yet to
be a large-scale, longitudinal examination of China's DNS filtering mechanism.
This lack of visibility is apparent as the number of censored domains and the
pool of IP addresses used by the GFW in forged DNS responses have been
reported differently by previous studies~\cite{Zittrain2003, Lowe2007a,
brown2010dns, Sparks:CCR2012, china:2014:dns:anonymous,
farnan2016cn.poisoning, Iris, Tripletcensors}. In particular, the number of
fake IPs observed in poisoned responses has been increasing from nine in
2010~\cite{brown2010dns}, 28 in 2011~\cite{Sparks:CCR2012}, 174 in
2014~\cite{china:2014:dns:anonymous}, to more than 1.5K
recently~\cite{Tripletcensors}. To that end, it is necessary to have a system
for continuous, long-term monitoring of the GFW's filtering policy that will
provide timely insights about its blocking behavior and assist censorship
detection and circumvention efforts.

In this work, we developed \sysname\ (\sectionref{sec:method}), a large-scale,
longitudinal measurement platform to shed light on DNS filtering by the GFW
and assess its impact on the global Internet. By building \sysname, our
primary goal is not only to answer the questions of \emph{(1) how many
censored domains are there} and \emph{(2) what are the forged IP addresses
used in fake DNS responses}, but also to assess \emph{(3) the impact of the
GFW's DNS censorship policy on the global Internet}, and ultimately design
\emph{(4) strategies to effectively detect and circumvent the GFW's DNS
censorship}.

Using \sysname, we tested a total of 534M distinct domains (averaging 411M
domains per day) and detected a total of 311K censored domains
(\sectionref{sec:censored_domains}). We then used the set of censored domains
to design a probing method that is able to reverse-engineer the actual
blocklist used by the GFW's DNS filter
(\sectionref{sec:reverse_engineer_blocklist}). Using this list, we observed
that 270K out of the 311K censored domains are censored as intended, whereas
the remaining 41K domains appear to be innocuous despite matching regular
expressions used by the GFW. Through our measurements, we discovered 1,781
IPv4 and 1,799 IPv6 addresses used by the GFW in forged DNS responses
(\sectionref{sec:forged_ips}). To the best of our knowledge, these are the
largest sets of censored domains and forged IP addresses ever discovered.

We also found evidence of geographic restrictions on Chinese domains, with the
GFW injecting DNS replies for domains based in China (e.g.,
\texttt{www.beian.gov.cn})  (\sectionref{sec:censorship_leakage}). While
previous studies attribute leakage of Chinese DNS censorship to cases where a
DNS resolver's network path transits through China's
network~\cite{brown2010dns, Sparks:CCR2012}, we found that geoblocking and
cases where censored domains have at least one authoritative name server
located in China are also a significant cause of pollution of external DNS
resolvers (\sectionref{sec:impact}).

Based on the observed censored domains (\sectionref{sec:censored_domains}) and
forged IP addresses (\sectionref{sec:forged_ips}), we propose strategies to
effectively detect poisoned DNS responses injected by the GFW
(\sectionref{sec:detection}). These techniques will not only help public DNS
resolvers and other DNS-related services to sanitize tainted records
(\sectionref{sec:detection}), but can also assist future development of
circumvention tools to bypass the GFW's DNS censorship
(\sectionref{sec:circumvention}).

%% file: background.tex
\section{Background}
\label{sec:background_motivation}

The Internet filtering infrastructure of China, allegedly designed in the late
90s under the Golden Shield project~\cite{Conoway2009, china-gfw}, is a system
used by the Chinese government to regulate the country's domestic Internet
access. The filtering system, commonly referred to as the Great
Firewall~\cite{Barme1997}, consists of middleboxes distributed across border
autonomous systems~\cite{crandall2007.conceptdoppler, xu:2011:china,
china:2014:dns:anonymous}, which are controlled in a centralized
fashion~\cite{Barme1997, Zittrain2003, Conoway2009, Deibert10chinacyberspace}.
There are several filtering modules developed to control the free flow of
information at different layers of the network stack, including TCP/IP packet
filtering~\cite{Clayton2006IgnoringTG, Park2010EmpiricalSO, Winter2012,
Nobori2014, Ensafi2015, Arun:foci18} and application-level keyword-based
blocking~\cite{Barme1997, Zittrain2003, Clayton2006IgnoringTG,
Rambert2021ChineseWO}. However, we focus our discussion on the DNS poisoning
aspect of the GFW which is relevant to our study. 

Unencrypted and unauthenticated DNS traffic is widely targeted by censorship
systems to interrupt communications between users and remote destinations
where censored content or services are hosted~\cite{Sparks:CCR2012, holdonDNS,
Satellite, Iris, ICLab:SP20}. Exploiting DNS insecurity, the GFW is designed
as an on-path/man-on-the-side (MotS) system which takes advantage of UDP-based
DNS resolution to inject fake responses when censored domains are detected in
users' DNS queries.

More specifically, when the GFW detects a DNS query for a censored domain, it
will forge a response with an incorrect DNS record towards the client. Some
specific domains (e.g., \texttt{google.sm}) can trigger the GFW to emit up to
three forged responses~\cite{Tripletcensors}. As an on-path system, the GFW
cannot modify or drop the legitimate response returned by the blocked domain's
authoritative name server or the public resolver chosen by the client.
However, since the GFW is usually closer (in terms of physical/network
distance) to the client, the injected response will usually arrive ahead of
the legitimate one (\sectionref{sec:evaluation}), thus being accepted by the
client who is now unable to access the domain.

%% file: method.tex
\section{\sysname\ Design}
\label{sec:method}

We designed \sysname\ according to the following requirements: (1) the
platform should be able to discover as many censored domains and forged IPs as
possible in a timely manner. More specifically, \sysname\ should be able to
obtain and test new domain names \emph{as they appear on the Internet}. (2) As
a longitudinal measurement platform, once a domain is discovered to be
censored, \sysname\ should continuously keep track of its blocking status to
determine whether the domain stays censored or becomes unblocked at some point
in the future. (3) By measuring many domains with sufficient frequency,
\sysname\ is expected to provide us with a good view into the pool of forged
IPs used by the GFW.

\subsection{Test Domains}

We are interested in the timely discovery of as many censored domains as
possible because we hypothesize that the GFW does not block just well-known
domains (e.g., \texttt{facebook.com}, \texttt{twitter.com},
\texttt{tumblr.com}) but also less popular or even unranked ones that are of
interest to smaller groups of at-risk people (e.g., political dissidents,
minority ethnic groups), who are often suppressed by local
authorities~\cite{china-uighurs-suppress}. Therefore, we opt to curate our
test list from top-level domain (TLD) zone files obtained from various
sources, including Verisign~\cite{Verisign} and the Centralized Zone Data
Service operated by ICANN~\cite{ICANN}, which we refresh on a daily basis.
Using zone files not only provides us with a good coverage of domain names on
the Internet, but also helps us to fulfill the first design goal of \sysname,
which is the capability to test new domains as they appear on the Internet.

Since TLD zone files contain only second-level domains (SLDs), they do not
allow us to observe cases in which the GFW censors subdomains of these SLDs.
As we show later, many subdomains (e.g., \texttt{scratch.mit.edu},
\texttt{nsarchive.gwu.edu}, \texttt{cs.colorado.edu}) are censored but their
SLDs (e.g., \texttt{mit.edu}, \texttt{gwu.edu}, \texttt{colorado.edu}) are
not. We complement our test list by including domains from the Citizen Lab
test lists (CLTL)~\cite{CLBL}, the Tranco list~\cite{LePochat2019}, and the
Common Crawl project~\cite{CommonCrawl}. Between April and December 2020, we
tested a total of 534M domains from 1.5K TLDs, with an average of 411M domains
daily tested.

\subsection{Measurement Approach} 

\begin{figure}[t]
    \centering
    \includegraphics[width=0.9\columnwidth]{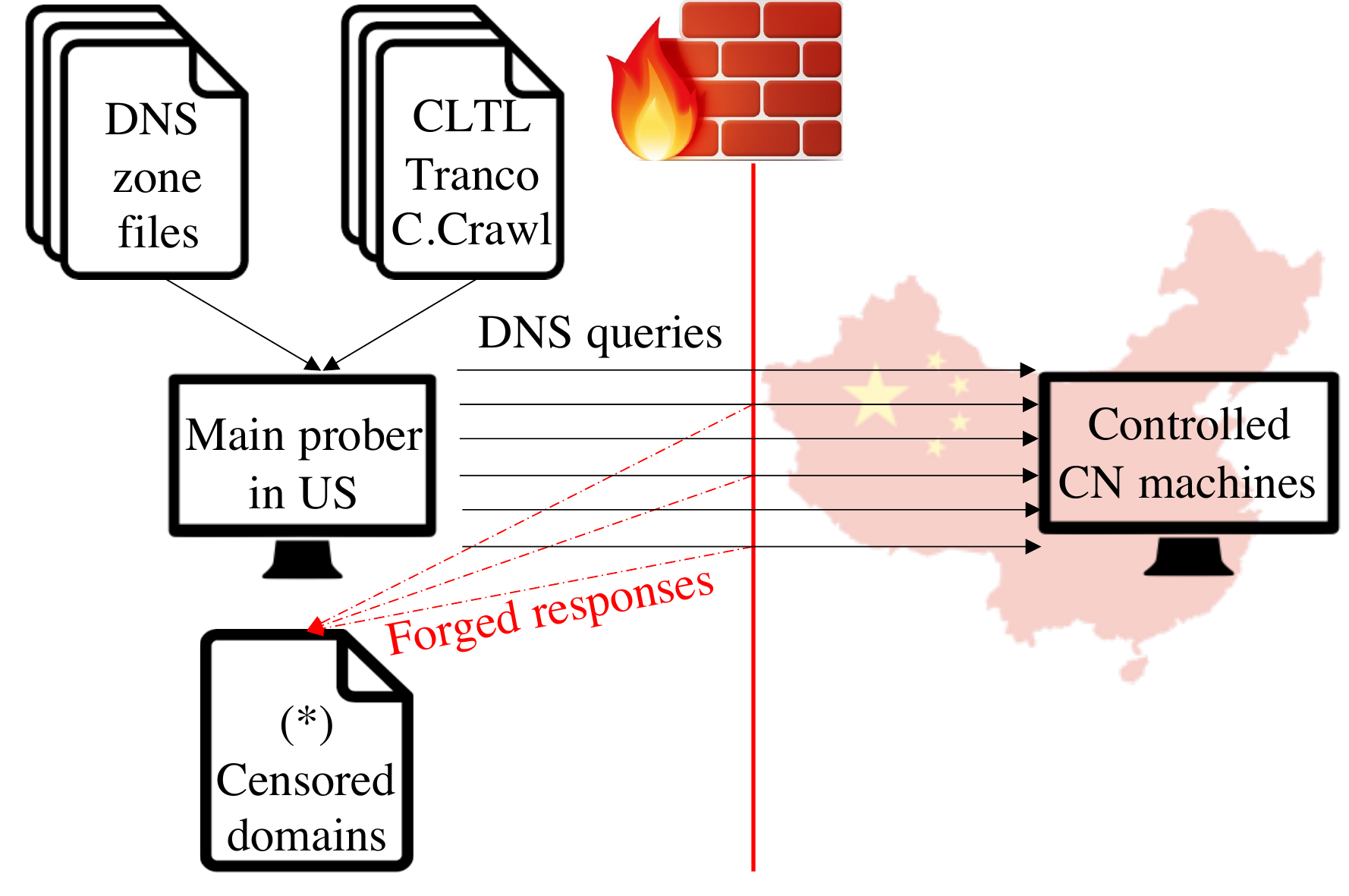}
    \caption{Probing the GFW's DNS poisoning from outside.}
    \label{fig:outside_in_probe}
    \vspace{-1.08em}
\end{figure}

When filtering DNS traffic, the GFW does not consider the direction of request
packets. As a result, even DNS queries originating from outside the country
can trigger the GFW if they contain a censored domain, making this behavior a
popular topic for measurement studies~\cite{brown2010dns, Sparks:CCR2012,
china:2014:dns:anonymous, Tripletcensors}. Based on the observation of this
filtering policy, we design \sysname\ to probe the GFW from outside of China
to discover censored domains and verify their blockage again from our
controlled machines located in China to validate our findings.

Prior work has shown that the GFW does not filter DNS traffic on ports other
than the standard port 53~\cite{Lowe2007a, Tripletcensors}, we thus design our
probe queries using this standard destination port number. We observe that for
major UDP-based DNS query types (e.g., \texttt {A}, \texttt{CNAME},
\texttt{MX}, \texttt{NS}, \texttt{TXT}), the GFW injects the forged responses
with an IPv4 for type \texttt{A} queries and a bogus IPv6 for type
\texttt{AAAA} queries. In some rare cases, injections of forged static CNAME
records are also observed for a small number of censored domains
(\sectionref{sec:forged_ips_grouping}).

For TCP-based queries that carry censored domains, RST packets are injected
instead of DNS responses~\cite{Wang2017YourSI}. Since UDP is the default
protocol for DNS in most operating systems, we choose to probe the GFW with
UDP-based queries. While using both TCP-based and UDP-based queries would
still allow us to detect censored domains, we opt to use UDP-based queries
because they also allow us to (1) collect the forged IPs used in the injected
DNS responses, and (2) conduct our measurement at scale, which would be
otherwise more challenging to achieve because a TCP-based measurement at the
same scale would require more computing and network resources to handle
stateful network connections.

As shown in Figure~\ref{fig:outside_in_probe}, \sysname's main prober is a
machine located in an academic network in the United States, where DNS
censorship is not anticipated. \texttt{A} and \texttt{AAAA} DNS queries for
the test domains are sent towards two hosts in China, which are under our
control and do not have any DNS resolution capabilities. Therefore, any DNS
responses returned to the main prober come from the GFW.

While prior studies have confirmed the centralized blocking policy of the
GFW~\cite{Barme1997, Conoway2009, Deibert10chinacyberspace}, to make sure this
behavior is still consistent and to detect any future changes, the two hosts
in China are located in two different autonomous systems (ASes). From our
measurement results, we confirm that the DNS blocking policy continues to be
centralized, with the same censored domains detected via the two probing
paths.

\begin{figure}[t]
    \centering
    \includegraphics[width=0.78\columnwidth]{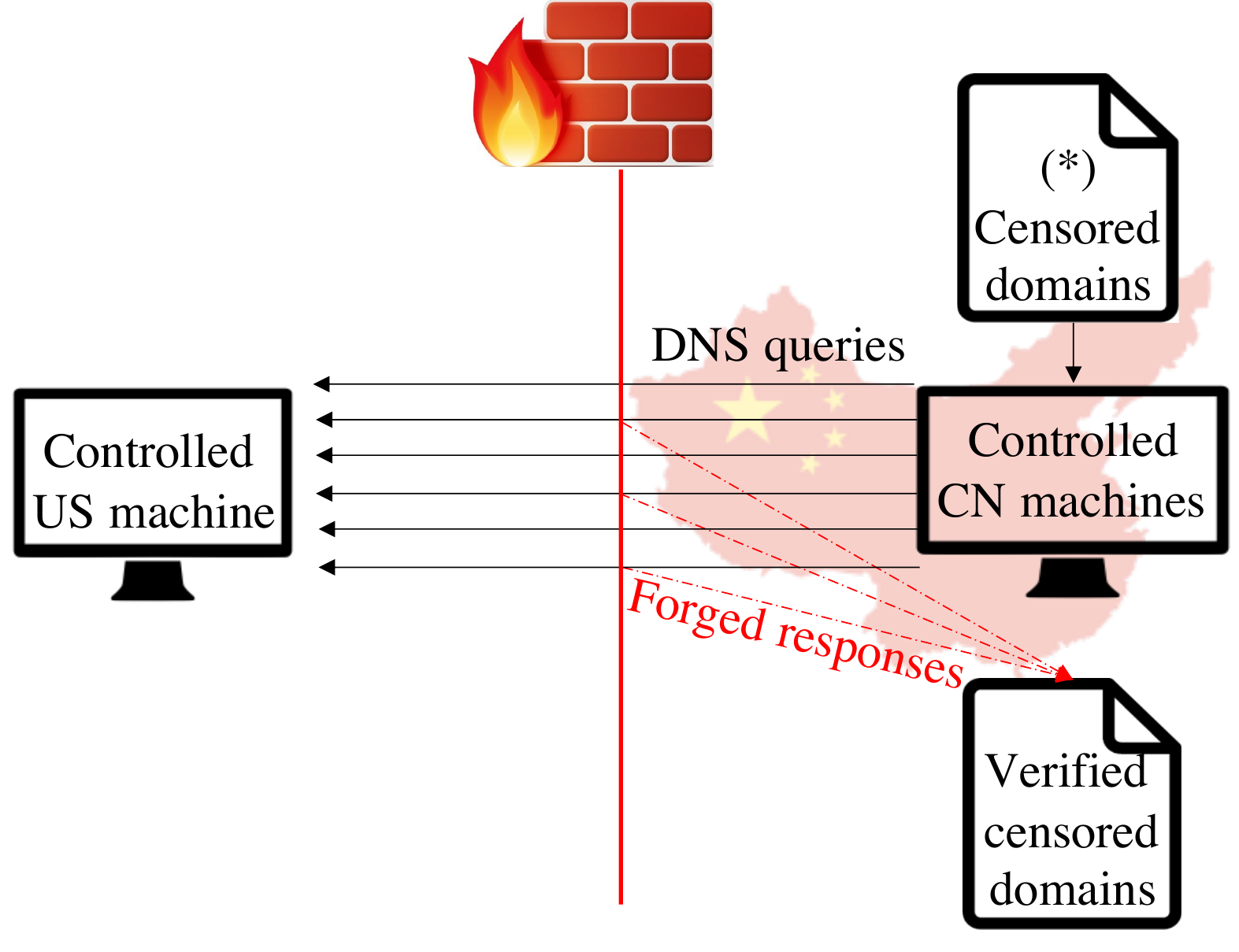}
    \caption{Verifying poisoned domains from inside the GFW.}
    \label{fig:inside_out_probe}
    \vspace{-1.08em}
    \end{figure}

After the main prober completes each probing batch, detected censored
domains are transferred to the Chinese hosts and probed again from inside
China towards our control machine, as shown in
Figure~\ref{fig:inside_out_probe}. This way, we can verify that censored
domains discovered by our prober in the US are also censored inside China.

Since \sysname\ is designed to probe using UDP, which is a stateless and
unreliable protocol, packets may get lost due to factors that are not under
our control (e.g., network congestion). Moreover, previous studies have
reported that the GFW sometimes fails to block access when it is under heavy
load~\cite{ensafi2015analyzing, Tripletcensors}. Therefore, to minimize the
impact of these factors on our data collection, \sysname\ tests each domain at
least three times a day.

For this paper, we use data collected during the last nine months of 2020,
from April to December. As of this writing, \sysname\ is still running and
collecting data every day. The data collected will be made available to the
public on a daily basis through a dedicated web service.

%% file: censored_domains.tex
\section{Censored Domains}
\label{sec:censored_domains}

\begin{figure}[t]
    \centering
    \includegraphics[width=0.8\columnwidth]{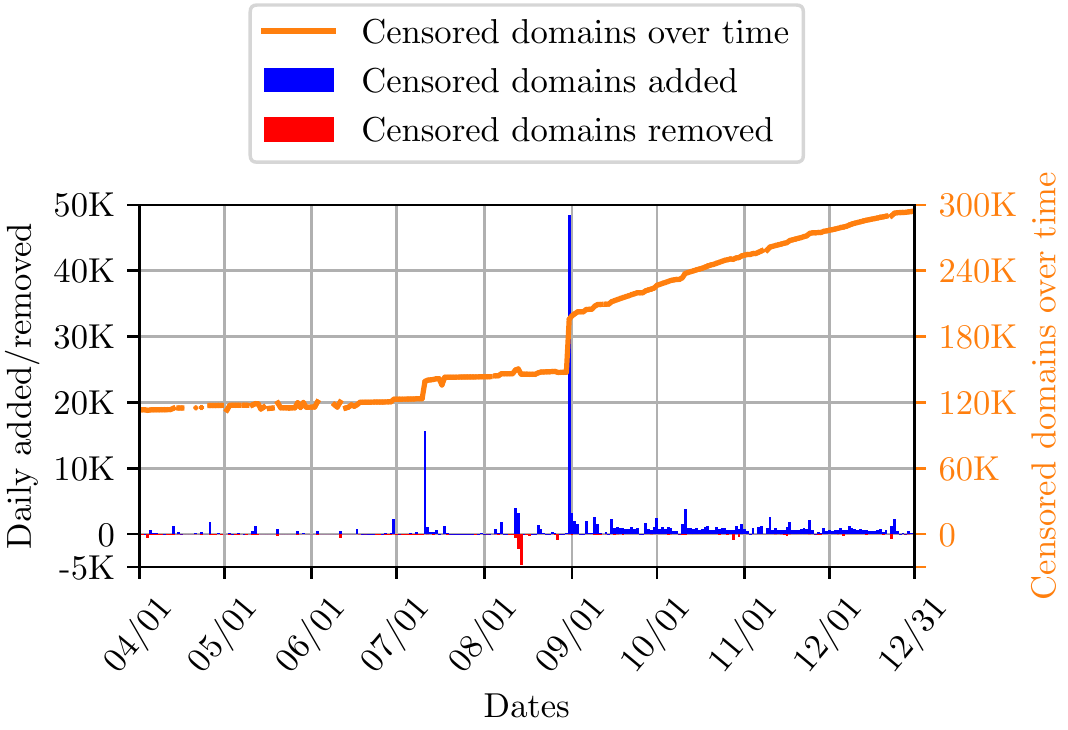}
    \caption{Cumulative censored domains discovered over time and daily
    added/removed censored domains.}
    \label{fig:censored_domains_churn}
    \vspace{-1.08em}
\end{figure}

Over the nine months of our study, we tested a total of 534M distinct domains,
finding 311K domains triggering the GFW's DNS censoring capability.
Figure~\ref{fig:censored_domains_churn} summarizes the cumulative number of
censored domains over time, as well as the number of domains added and removed
from the set of censored domains each day. We note a sharp increase in domains
on August 31st because of the addition of more than 30K subdomains from the
previously censored namespaces (e.g., \texttt{*.googlevideo.com},
\texttt{*.appspot.com}) to our test domains. In this section, we describe our
technique for identifying the specific strings that trigger GFW's DNS
censorship (\sectionref{sec:reverse_engineer_blocklist}). We use this
technique to remove unrelated domains that match the blocking rules
(``overblocked'' domains) and then characterize domains censored by the GFW in
Section~\ref{sec:censored_domains_taxonomy}.

\subsection{Identifying Blocking Rules}
\label{sec:reverse_engineer_blocklist}

When considering the domains filtered by the GFW, there are many
with common second-level and top-level domains
(e.g., numerous blocked domains of
the form \texttt{*.blogspot.com} or \texttt{*.tumblr.com}). This observation
led us to develop a clustering method for domains that are
blocked based on the same underlying rule. For example, if
\texttt{subdomain.example.com} and all subdomains of \texttt{example.com} are
blocked, we consider \texttt{example.com} as the blocked domain. We note
that when a subdomain is blocked, the covering
domains may not be blocked
(e.g., \texttt{cs.colorado.edu} is blocked,
whereas \texttt{colorado.edu} is not
(\sectionref{sec:censored_domains_taxonomy})).

Inspired by a previous study of GFW's DNS
censorship~\cite{china:2014:dns:anonymous}, we use the following technique to
identify the strings that trigger blocking (i.e., the most general string such
that all domains containing this string are blocked). For a given domain, we
test the following permutations of each censored domain and random strings: 

\begin{itemize}[noitemsep,topsep=0pt]
    \item Rule 0 \texttt{censored\_domain}
    \item Rule 1 \texttt{censored\_domain\emph{\{.rnd\_str\}}}
    \item Rule 2 \texttt{censored\_domain\emph{\{rnd\_str\}}}
    \item Rule 3 \texttt{\emph{\{rnd\_str.\}}censored\_domain}
    \item Rule 4 \texttt{\emph{\{rnd\_str\}}censored\_domain}
    \item Rule 5 \texttt{\emph{\{rnd\_str.\}}censored\_domain\emph{\{.rnd\_str\}}}
    \item Rule 6 \texttt{\emph{\{rnd\_str.\}}censored\_domain\emph{\{rnd\_str\}}}
    \item Rule 7 \texttt{\emph{\{rnd\_str\}}censored\_domain\emph{\{.rnd\_str\}}}
    \item Rule 8 \texttt{\emph{\{rnd\_str\}}censored\_domain\emph{\{rnd\_str\}}}
\end{itemize}

 Among these rules, only Rules~1 and~3 are correct forms of a domain with a
 different top-level domain (Rule~1) or subdomain (Rule~3). In contrast, the
 rest represents unrelated (or non-existent) domains that happen to contain
 the censored domain string. We refer to censored domains that are grouped
 with a shorter domain string via rules other than Rules~1 or~3 as being
 \emph{overblocked}, because they are not subdomains of the shorter domain,
 but are actually unrelated domains that are textually similar (e.g., the
 censored domain \texttt{mentorproject.org} contains the shorter domain string
 \texttt{torproject.org} that actually triggers censorship).

Using these rules to generate domains and testing them with \sysname, we
identify the most general form of each censored domain that triggers censorship.
We refer to these shortest censored domains as the ``base domain'' from which
the blocking rule is generated. We discovered a total of 138.7K base domains
from the set of 311K censored domains.

\begin{figure}[t]
    \centering
   \includegraphics[width=0.8\columnwidth]{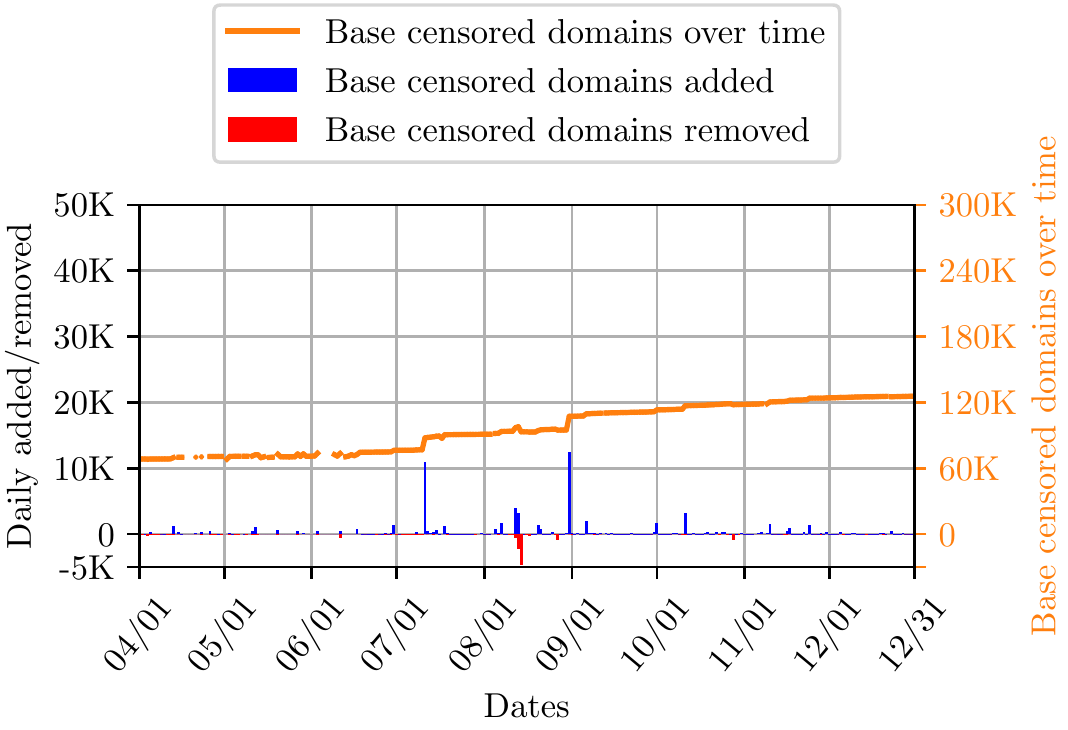}
   \caption{Cumulative base censored domains discovered over time and daily
   added/removed base censored domains.}
   \label{fig:base_censored_domains_churn}
   \vspace{-1.08em}
\end{figure}

Considering base domains allows us to observe growth in the underlying
blocking rules as opposed to the raw number of domains. We also observe fewer
new base domains over time and avoid sudden jumps in censored domains when
large numbers of subdomains of an existing base domain are observed.
Figure~\ref{fig:base_censored_domains_churn} shows the cumulative number of
base domains discovered over the nine-month period and the daily addition and
removal of these domains. As of December 31st, 126K base domains
are still being censored.

Of 138.7K base domains, 11.8K are censored independently (Rule 0). In other
words, these domains are censored as they are, but do not trigger GFW's DNS
censorship when concatenated with random strings. However, in an ascending
order of severity, we find that 4, 113.8K, 10.9K, 1.4K, and 696 distinct base
domains are blocked under Rules 2, 3, 4, 6, and 8, respectively. There are no
domains for Rules 1, 5, and 7, since domains blocked under these rules are
already covered by other more general rules. While the vast majority of base
censored domains fall under Rule 3, there are more than 13K base domains
blocked under other rules, causing unrelated domains to be overblocked.

We utilize the base domains to identify cases of overblocking, where an
unrelated domain matches a more general censored domain string. Specifically,
we consider domains that match a base domain, but are not subdomains of the
base domain, as being overblocked. This is because these domains are unrelated
to the base domain despite being textually similar. With this definition, we
find that 41K of the 331K censored domains are overblocked. The top three base
domains that cause the most overblocking are \texttt{919.com},
\texttt{jetos.com}, and \texttt{33a.com}. These three domains are responsible
for a total of 15K unrelated domains being blocked because they end with one
of these three base domains (and are not subdomains of them).
Table~\ref{table:innocuous_censored_domains} in
Appendix~\ref{sec:extreme_blocking rules} provides more details on the base
domains responsible for the most overblocking. Domain owners may consider
refraining from registering domain names containing these base domains to
avoid them being inadvertently blocked by the GFW.

\subsection{Characterizing Censored Domains}
\label{sec:censored_domains_taxonomy}

We now characterize the 138.7K base domains identified
in~\sectionref{sec:reverse_engineer_blocklist}. We focus on these base domains
to avoid the impact of domains with numerous blocked subdomains on our
results. Focusing on base domains also allows us to avoid analyzing innocuous
domains that are overblocked based on our previous analysis.

\begin{figure}[t]
    \centering
    \includegraphics[width=0.8\columnwidth]{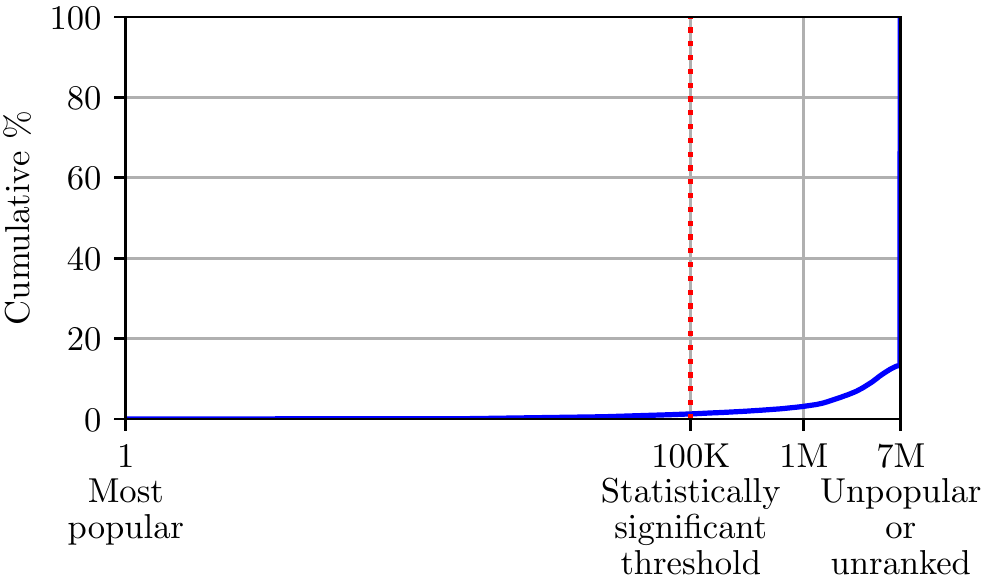}
    \caption{CDF of the popularity ranking for base censored domains (in log
    scale).}
    \label{fig:base_censored_domain_popularity}
    \vspace{-1.08em}
\end{figure}

\myparab{Popularity of censored domains.}
We find that most domains blocked by the GFW are unpopular and do
not appear on lists of most popular websites. We use the rankings
provided by the Tranco list~\cite{LePochat2019}, which combines
four top lists (Alexa~\cite{alexa}, Majestic~\cite{majestic},
Umbrella~\cite{cisco_umbrella}, and Quantcast~\cite{quantcast}) in a way that
makes it more stable and robust against malicious
manipulations~\cite{LePochat:CSET2019}. The daily Tranco list
contains about 7M domains ranked by the Dowdall rule~\cite{Fraenkel2014}.

Figure~\ref{fig:base_censored_domain_popularity} shows the CDF of the
popularity ranking for the 138.7K blocked base domains. Only 1.3\% of them
are among the top 100K most popular domains, which
is the statistically significant threshold of the popularity ranking as
suggested by both top-list providers and previous studies~\cite{alexa.qa2,
Rweyemamu2019}. Even when considering all domains ranked by the Tranco list,
only 13.3\% of the base censored domains fall within the list's ranking range,
while the remaining are unranked. This
finding highlights the importance of \sysname's use of TLD zone files to
enumerate the set of potentially censored domains.

\begin{figure}[t]
    \centering
    \includegraphics[width=0.8\columnwidth]{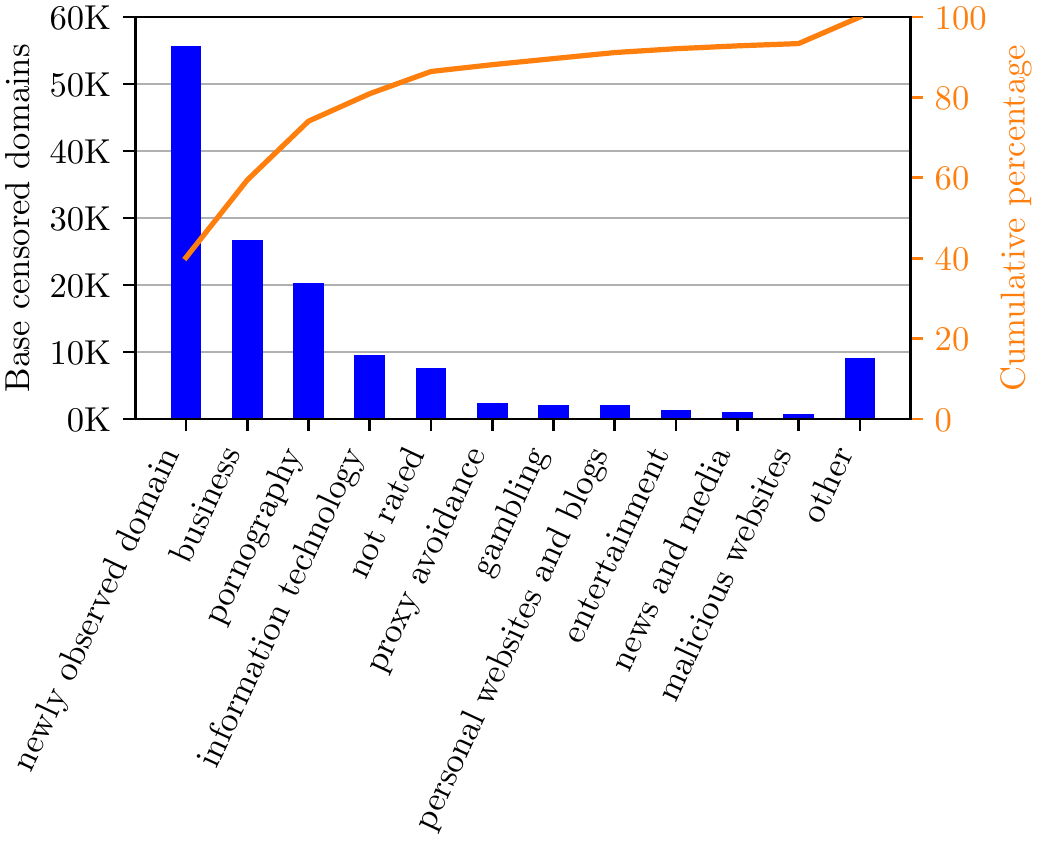}
    \vspace{-0.8em}
    \caption{Top ten categories of domains censored by the GFW.}
    \label{fig:top_10_censored_categories}
    \vspace{-1.08em}
\end{figure}

\myparab{Types of censored content.} For domain categorization, we use a
service provided by FortiGuard~\cite{FortiGuard}, which has also been used by
other censorship measurement studies~\cite{Tripletcensors, ICLab:SP20,
Raman2020CensoredPlanet}, to make our analysis comparable.
Figure~\ref{fig:top_10_censored_categories} shows the top-ten domain categories 
censored by the GFW. We find that nearly half of the domains we
observe are not currently categorized by FortiGuard, with 40\% categorized
as \emph{``newly observed domain,''}
and 5.5\% categorized
as \emph{``not rated.''} This is a result of the large number of domains in
our dataset, many of which may not be currently active
(\sectionref{sec:trueRR}).

Apart from the \emph{``newly observed domain''} and \emph{``not rated''}
categories, we find that \emph{``business,''} \emph{``pornography,''} and
\emph{``information technology''} are within the top-five dominant categories.
This finding is different from the results reported by the most recent related
work to ours~\cite{Tripletcensors}, which observed \emph{``proxy avoidance''}
and \emph{``personal websites and blogs''} as the most blocked categories.
This difference stems from the counting process used in~\cite{Tripletcensors},
which does not aggregate subdomains, while their test list is a fixed snapshot
of 1M domains from the Alexa list, which contains many subdomains of
\texttt{*.tumblr.com} and \texttt{*.blogspot.com}.

\myparab{COVID-19 related domains.} On December 19th, 2020, the New York Times
reported that the Chinese Government issued instructions for suppressing the
free flow of information related to the COVID-19
pandemic~\cite{CN-Covid-Censorship}.  \sysname\ has detected numerous domains
related to COVID-19 being censored by the GFW through DNS tampering, including
\texttt{covid19classaction.it}, \texttt{covid19song.info}
\texttt{covidcon.org}, \texttt{ccpcoronavirus.com}, \texttt{covidhaber.net},
and \texttt{covid-19truth.info}.

While most censored domains are discovered to be blocked soon after they
appear in our set of test domains, we found that there was some delay in
blocking \texttt{ccpcoronavirus.com}, \texttt{covidhaber.net}, and
\texttt{covid-19truth.info}. Specifically, \texttt{ccpcoronavirus.com} and
\texttt{covidhaber.net} first appeared on our test lists in April but are not
blocked until July and September, respectively. Similarly,
\texttt{covid-19truth.info} appeared in our dataset in September but was not
censored until October. The large difference in the time the GFW
takes to censor different domains shows that the blocklist is likely to be
curated by both automated tools and manual efforts.

\myparab{Educational domains.} In 2002, Zittrain et al.~\cite{Zittrain2003}
reported DNS-based filtering of several institutions of higher education in
the US, including \texttt{mit.edu}, \texttt{umich.edu}, and \texttt{gwu.edu}.
While \emph{``education''} is not one of the top censored categories, we find
numerous blocked education-related domains, including
\texttt{armstrong.edu}, \texttt{brookings.edu}, \texttt{citizenlab.ca},
\texttt{feitian.edu}, \texttt{languagelog.ldc.upenn.edu}, \texttt{pori.hk},
\texttt{soas.ac.uk}, \texttt{scratch.mit.edu}, and \texttt{cs.colorado.edu}.

Although censorship against some of these domains is not
surprising, since they belong to institutions well-known for conducting
political science research and may host content deemed as unwanted, we are
puzzled by the blocking of \texttt{cs.colorado.edu}. While the University of
Colorado's computer science department is not currently using this domain to
host their homepage, the blocking of this domain and its entire namespace
\texttt{*.cs.colorado.edu} would prevent students in China from accessing
other department resources (e.g., \texttt{moodle.cs.colorado.edu}). This is
another evidence of the overblocking policy of the GFW, especially during the
difficult time of the COVID-19 pandemic when most students need to take
classes remotely.

%% file: forged_ips.tex
\section{Forged IP Addresses}
\label{sec:forged_ips}

The use of publicly routable IPs owned by foreign entities not only confuses
the impacted users and misleads their interpretation of the GFW's censorship,
but also hinders straightforward detection and
circumvention~\cite{greatfire2015}. Therefore, knowing the forged IPs and the
pattern in which they are injected (if any) is essential. In this section, we
analyze the IPs collected by \sysname\ to examine whether there exists any
specific injection pattern based on which we can develop strategies to
effectively detect and bypass the GFW's DNS censorship.

\subsection{Forged IP Addresses over Time}
\label{sec:forged_ips_over_time}

Extracting the forged IPs from all poisoned DNS responses captured by
\sysname, we find a total of 1,781 and 1,799 unique forged IPv4 and IPv6
addresses from poisoned type-A and type-AAAA responses, respectively. The
forged IPv4 addresses are mapped to multiple ASes owned by numerous
non-Chinese entities, including 783 (44\%) IPs of Facebook, 277 (15.6\%) IPs
of WZ Communications Inc., 200 (11.2\%) IPs of Twitter, and 180 (10.1\%) IPs
of Dropbox. On the other hand, all IPv6 addresses are bogus and belong to the
same subnet of the predefined Teredo prefix~\cite{rfc4380},
\texttt{2001::/32}. Therefore, we will focus our analysis on the forged IPv4
addresses hereafter because the pattern of IPv6 injection is obvious and thus
should be trivial to detect and circumvent.

\begin{figure}[t]
  \centering
  \includegraphics[width=0.8\columnwidth]{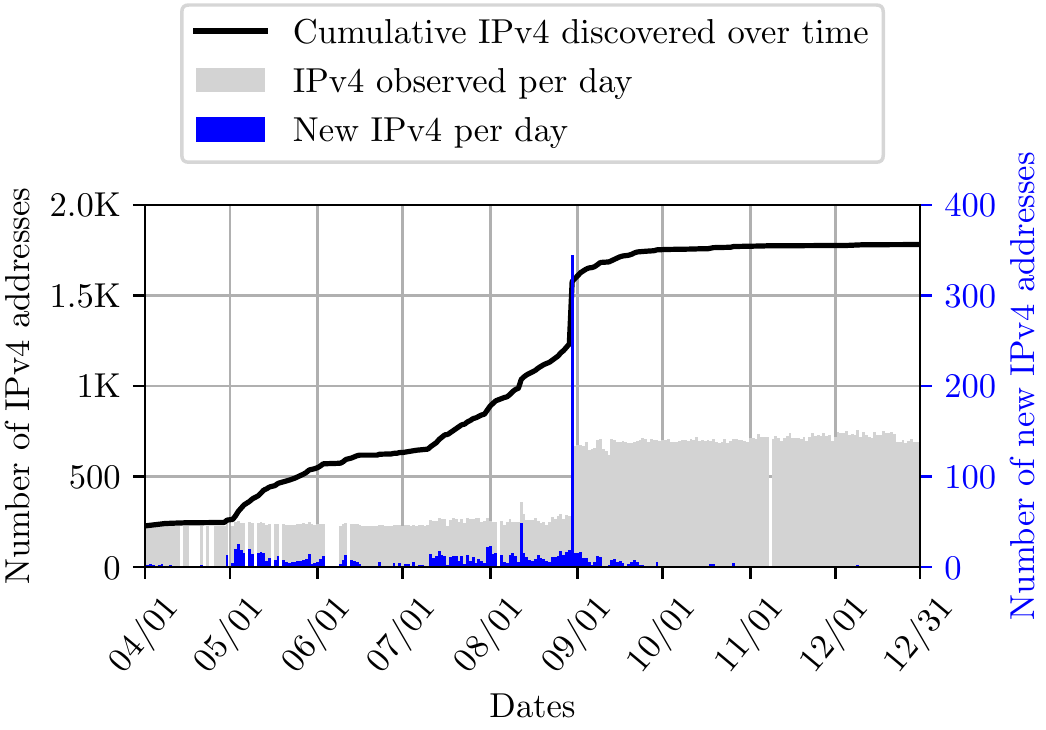}
  \caption{Number of forged IPv4 addresses detected over time by \sysname.}
  \label{fig:ipv4_overtime}
  \vspace{-1.08em}
\end{figure}

Figure~\ref{fig:ipv4_overtime} shows the number of unique IPv4 addresses that
\sysname\ has discovered over the measurement period considered in this paper.
The gray bar plot shows the number of unique IPs observed daily, and the blue
bar plot shows the number of new IPs that were not observed previously. We add
a second y-axis on the right side of the figure for better visibility of the
blue bars.

Our initially collected data overlaps with the data collected during the final
month of~\cite{Tripletcensors}, which is the most recent related work to our
study. During this period, our observation aligns with the result reported in
Figure 2 of~\cite{Tripletcensors}, i.e., the number of unique forged IPs is
about 200 with no new IPs detected. However, starting in May, \sysname\ began
to detect more forged IPs every day until September, with about 10--20 new IPs
added daily. These gradual daily additions, together with a significant
increase of more than 300 previously unobserved IPs at the end of August, have
brought the total number of forged IPs to more than 1.5K. The number of forged
IPs converges to 1.7K over the last four months of 2020.

Comparing the IPs observed by \sysname\ with the ones reported
in~\cite{Tripletcensors}, we find that all IPs observed
by~\cite{Tripletcensors} have been used again in poisoned DNS responses,
regardless of the major drop reported on November 23rd, 2019. In addition, we
find 188 new IPs that were not observed previously in~\cite{Tripletcensors}.
Given how close the timeline is between our work and~\cite{Tripletcensors},
this finding of the unpredictable fluctuation in the number of forged IPs
emphasizes the importance of having a large-scale longitudinal measurement
system to keep track of erratic changes in the GFW's blocking behavior.
Therefore, we are committed to keeping \sysname\ running as long as possible,
rather than just creating it as a one-off effort.

Prior reports~\cite{Barme1997, Conoway2009, Deibert10chinacyberspace}
and our detection of the same censored domains via two different network paths
(\sectionref{sec:method}) have confirmed the centralized blocking policy of
the GFW in terms of the domains being censored. Nevertheless, we are also interested in
investigating whether the forged IPs are consistent at different network
locations, because our ultimate goal is to collect as many forged IPs as
possible and demystify their injection pattern to assist us in developing
effective strategies for censorship detection and circumvention. Therefore, we
have also conducted an extra measurement by probing across different network
locations in China to confirm that the pool of forged IPs discovered by
\sysname\ is representative enough. More details of this measurement are
provided in Appendix~\ref{sec:ip_consistency}.

\subsection{Injection Frequency of Forged IPs}
 \label{sec:ip_frequency}

 \begin{figure}[t]
  \centering
  \includegraphics[width=0.8\columnwidth]{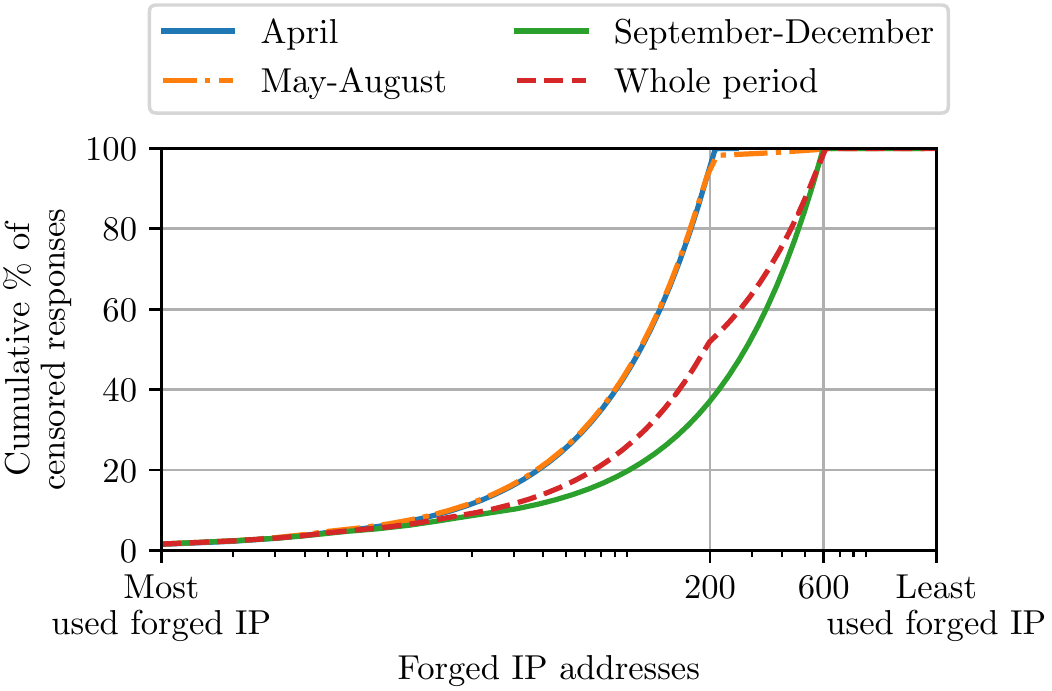}
  \caption{CDF of censored responses with respect to the injection frequency
  of forged IPv4 addresses detected by \sysname.}
  \label{fig:ipv4_percentiles_overtime}
  \vspace{-1.08em}
\end{figure}

Due to the erratic changes in the number of forged IPs over time, prior
studies have often concluded that forged IPs are injected randomly.
Through the longitudinal measurement conducted at scale, \sysname\ has tested and
detected a large enough number of censored domains and forged IPs that
allows us to provide more insights into this aspect. Analyzing the injection
frequency of each forged IP, we find that not all forged IPs are equally
injected in censored responses, i.e., their injection pattern is not
entirely random.

Figure~\ref{fig:ipv4_percentiles_overtime} shows the CDF of censored responses
with respect to the injection frequency of forged IPs observed in these
responses. The x-axis (in log scale) indicates the number of forged IPs,
sorted by their injection frequency. There are three periods during which the
cumulative number of forged IPs shows different patterns (i.e., April, May to
August, and September to December, as shown in
Figure~\ref{fig:ipv4_overtime}). Thus, we analyze the injection frequency of
these three periods independently and compare them with the injection
frequency of all forged IPs discovered over the whole period of our
measurement.

We can see that the forged IPs' injection frequencies are similar (almost
overlapping) between the April and May--August lines. In other words, although
the number of forged IPs increases from about 200 at the end of April to more
than 1.5K over the May--August period, the initial 200 forged IPs are still
responsible for 99\% of censored responses. On the other hand, the additional
1.3K new forged IPs discovered from May to August are in the long tail and
only used in 1\% of all censored responses. Similarly, even after the
remarkable increase to more than 1.7K forged IPs at the end of August, only
600 of them are frequently injected from September to December, occupying 99\%
of the censored responses. Finally, when looking at all the censored responses
and forged IPs discovered over the whole period, the 200 most frequently
injected forged IPs discovered in April are still responsible for more than
50\% of all censored responses, whereas only 600 (33.6\%) out of 1,781 forged
IPs are responsible for 99\% of all censored responses, the remaining 1.1K
forged IPs in the long tail are used in only 1\% of censored responses.

\subsection{Static and Dynamic Injections}
\label{sec:forged_ips_grouping}

\begin{table}[t]
  \centering
  \centering
  \caption{Groupings of censored domains with respect to different sets of
  forged IPs injected in their poisoned responses.}

  \resizebox{\columnwidth}{!}{
  \begin{tabular}{lrrl}
  \toprule
  \textbf{G} &\textbf{\# Domains} & \textbf{\# IPs} & \textbf{Forged IPs/CNAMEs} \\
  \midrule
  0 & 41        & 0  & cathayan.org, mijingui.com, upload.la, yy080.com\\
  \midrule
  1 & 12        & 1  & why.cc $\rightarrow$ 216.139.213.144\\
  \midrule
  2 & 7         & 1  & yumizi.com $\rightarrow$ 66.206.11.194\\

  \midrule
  3 & 57        & 1  & 46.38.24.209, 46.20.126.252, 61.54.28.6, 89.31.55.106\\
    &           &    & 122.218.101.190, 123.50.49.171, 173.201.216.6, 208.109.138.55\\
  \midrule
  4 & 3,295     & 3  & 4.36.66.178, 64.33.88.161, 203.161.230.171\\
  \midrule
  5 & 1,711     & 4  & 8.7.198.45, 59.24.3.173, 243.185.187.39, 203.98.7.65\\
  \midrule
  6 & 2,724     & 4  & 8.7.198.46, 59.24.3.174, 46.82.174.69, 93.46.8.90\\
  \midrule
  7 & 4         & 7  & 4.36.66.178, 64.33.88.161, 203.161.230.171, 59.24.3.174\\
    &           &    & 8.7.198.46, 46.82.174.69, 93.46.8.90 \\
  \midrule
  8 & 9         & 7  & 4.36.66.178, 64.33.88.161, 203.161.230.171, 8.7.198.45\\
    &           &    & 59.24.3.173, 243.185.187.39, 203.98.7.65\\
  \midrule
  9 & 4,551     & 10  & 23.89.5.60, 49.2.123.56, 54.76.135.1, 77.4.7.92\\
    &           &     & 118.5.49.6, 188.5.4.96, 189.163.17.5, 197.4.4.12\\
    &           &     & 249.129.46.48, 253.157.14.165\\
  \midrule
  10& remaining &$>$560 & [Omitted due to the large number of forged IPs]\\
    & $\sim300K$&       & Supplementary data will be made publicly available \\
    & domains   &       & and updated on a daily basis.\\

    \bottomrule
  \end{tabular}}
  \label{tab:ip_grouping}
  \vspace{-1.08em}
\end{table}

One of the GFW behaviors is injecting different sets of forged IPs for
different groups of censored domains. This behavior was first reported
in~\cite{Tripletcensors}, where the authors identify a total of six groups of
censored domains that are poisoned with different sets of forged IPs. From
data collected by \sysname, we have discovered a total of 11 groups shown in
Table~\ref{tab:ip_grouping}. Comparing these groups with those reported
in~\cite{Tripletcensors}, we find five similar groups that have the same set
of forged IPs/CNAMEs, including Groups 0, 4, 5, 6, and 9. Understandably, we
discover more groups because our test list covers far more domains compared
to~\cite{Tripletcensors}, where a fixed Alexa top list of only 1M domains was
used for the whole measurement period.

An instance of forged response containing a CNAME was reported
in~\cite{Tripletcensors} but excluded from the analysis since it did not seem
to be prevalent. However, with a larger dataset, we find that the injection of
CNAME in forged responses can happen in three different groups of censored
domains, triggering the GFW to inject six different CNAME answers. As depicted
in Table~\ref{tab:ip_grouping}, there are 41 censored domains that can trigger
the injection of \emph{either one of the four} CNAMEs listed. Domains in
Groups 1 and 2 can trigger a CNAME injection, accompanied by an IP in the
forged response. Note that these two IPs are not the actual IPs of the two
CNAMEs. Similarly, there are eight distinct subgroups of domains within Group
3 that can constantly trigger \emph{either one of the eight} forged IP listed.
For example, \texttt{qcc.com.tw} will always trigger a forged response of
\texttt{89.31.55.106}. The same pattern applies in other Groups from 4 to 9,
i.e., resolving domains within these groups will always trigger the GFW to
inject one of the forged IPs listed on the 4th column. The remaining of about
300K censored domains are grouped together since they trigger the GFW to
dynamically inject a much larger number of more than 560 different forged IPs.

Revealing these injection patterns for different groups of censored domains is
crucial for developing an effective strategy to detect and circumvent the
GFW's DNS censorship (\sectionref{sec:censorship_leakage}). Especially,
knowing whether a censored domain belongs to one of the static groups (Groups
0 to 9) or the dynamic group (Group 10) is necessary to avoid misclassifying
consistent forged responses as ``legitimate''
(\sectionref{sec:circumvention}).

%% file: censorship_leakage.tex
\section{Censorship Leakage and Detection}
\label{sec:censorship_leakage}

The GFW's bidirectional DNS filtering behavior has been reported as the cause
of poisoned DNS responses being cached by public DNS resolvers outside China,
when DNS resolution paths unavoidably have to transit via China's
network~\cite{Sparks:CCR2012, hoang:2019:measuringI2P}. However, in this
section, we show that DNS poisoning against many domains whose authoritative
name servers are located in China is another primary reason why poisoned DNS
records have tainted many public DNS resolvers around the world. We then show
how the datasets of censored domains and forged IPs discovered by \sysname\
can help with detecting and sanitizing poisoned resource records from public
DNS resolvers' cache.

\subsection{Geoblocking of China-based Domains}
\label{sec:impact}

On August 8th, 2020, \sysname\ detected the blockage of
\texttt{www.beian.gov.cn}, which is managed by the Chinese Ministry of
Industry and Information Technology. This service allows website owners to
obtain and verify their website's Internet Content Provider (ICP) license,
which is obligated to legally operate their site in China. This domain has two
authoritative name servers, \texttt{dns7.hichina.com} and
\texttt{dns8.hichina.com}, which are hosted on 16 different IPs. However,
checking against the latest MaxMind dataset~\cite{MaxMind}, we find that all
of these IPs are located inside China. Consequently, the DNS censorship
against this domain by the GFW will cause DNS queries issued from outside
China to be poisoned since all resolution paths from outside China will have
to cross the GFW.

We initially attributed this blockage to an error or a misconfiguration
because previous works have sometimes noticed intermittent failures in the
GFW~\cite{ensafi2015analyzing, Tripletcensors}. Furthermore, no prior studies
have ever found such a strange blocking behavior---the GFW of China censors a
Chinese government website. However, at the time of composing this paper, we
are still observing \texttt{www.beian.gov.cn} being censored by the GFW,
almost half a year since its first detection. Hence, this is a clear case of
geoblocking because we can still visit this domain normally from our
controlled machines located inside China. To the best of our knowledge, ours
is the first academic research to document this geoblocking behavior of the
GFW.

Note that this geoblocking is a result of the GFW's DNS censorship, which is
not the same as geoblocking enforced at the server
side~\cite{McDonald2018403FA}. Geoblocking of China-based websites has been
noticed previously but is enforced by their website owners. For instance,
political researchers have been using \texttt{https://www.tianyancha.com/} to
investigate the ownership of Chinese companies, but since 2019, this website
blocks visitors from non-Chinese IPs and shows a clear message for the reason
of denying access.

\begin{figure}[t]
  \centering
  \includegraphics[width=0.6\columnwidth,height=0.4\columnwidth]{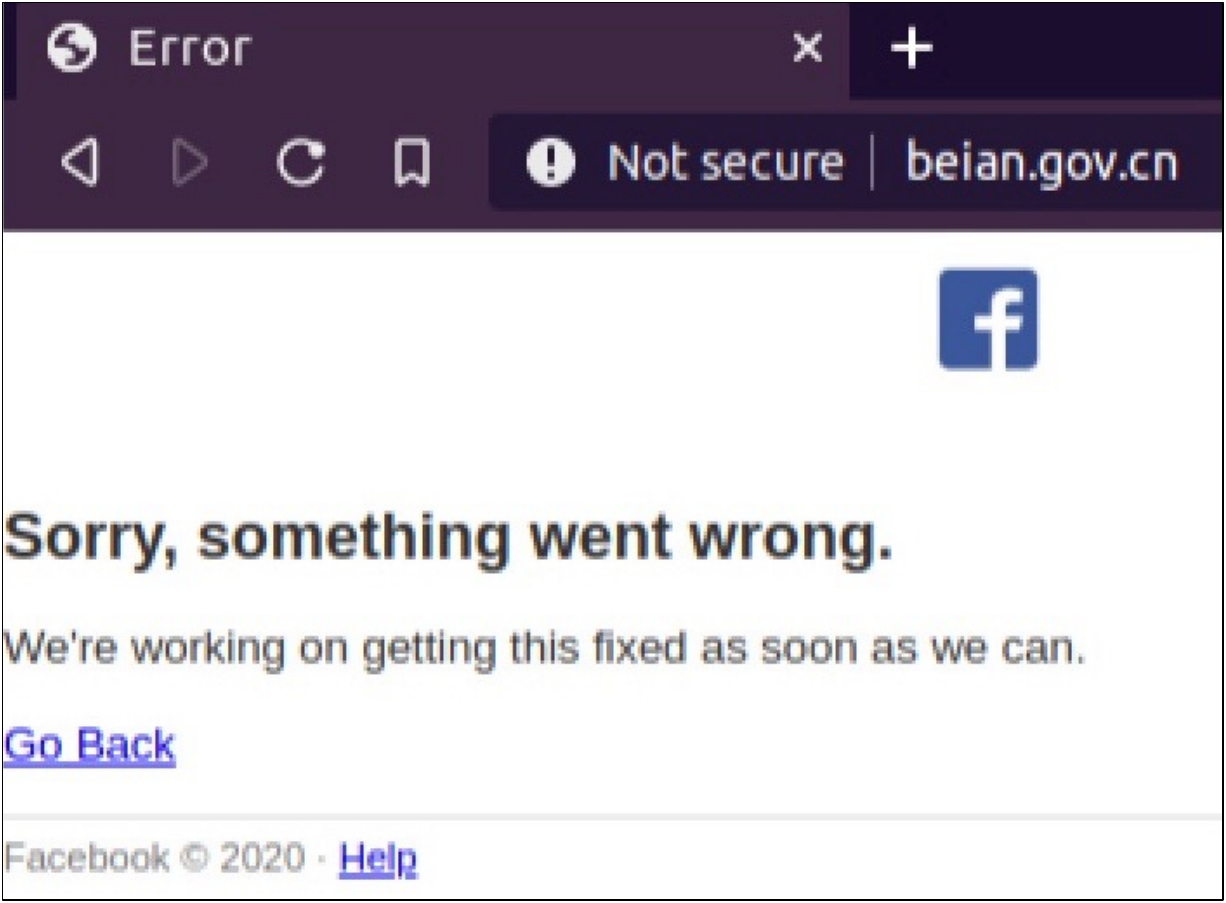}
  \caption{Visit to a domain geoblocked by the GFW ends up with an error page from Facebook.}
  \label{fig:facebook_error_page}
  \vspace{-1.08em}
\end{figure}

The GFW's blocking of China-based domains using bidirectional DNS filtering in
combination with the use of forged IPs owned by non-Chinese entities impacts
not only Internet users in China, but also users from around the world. For
instance, upon visiting the aforementioned geoblocked domain from a
non-censored network outside China, we end up with an error page served from
Facebook, as shown in Figure~\ref{fig:facebook_error_page}.

Most ordinary Internet users would not know the underlying reason why their
visit to a given China-based domain (e.g., \texttt{www.beian.gov.cn}) that is
clearly unrelated to Facebook would end up with an error page from Facebook.
The fact that the GFW frequently changes the forged IPs used in fake DNS
responses (\sectionref{sec:forged_ips}) would cause even more confusion to the
affected users. Depending on which fake IP is injected in the spoofed
response, users may encounter a different error page from
Figure~\ref{fig:facebook_error_page}. Even more confusing, the visit to this
domain from outside China will intermittently succeed because the poisoned
responses injected by the GFW sometimes fail to arrive ahead of the legitimate
one (\sectionref{sec:circumvention}).

At the server side of the forged IPs being used for injecting poisoned
responses, their operators would also be puzzled as to why many HTTP requests
are sent to their servers, asking for hostnames they do not serve. For the
above example, an error log at a Facebook server will show that someone was
trying to visit \texttt{www.beian.gov.cn} on a Facebook IP, which obviously
does not serve any content for that domain, thus the returned error page. As
we do not have access to the error logs of Facebook and other organizations
whose IPs are used for injecting poisoned DNS responses by the GFW, we cannot
quantify the actual cost (e.g., the overhead of serving unsolicited
connections, error pages) of such an abusive DNS redirection behavior.
However, given the large number of more than 311K censored domains discovered
(\sectionref{sec:censored_domains}) and only a small pool of forged IPs being
used (\sectionref{sec:forged_ips}), we believe that the GFW's injection policy
would cost these affected organizations a non-negligible overhead on their
servers. Past reports have shown that this abusive design of the GFW can lead
to resource exhaustion attacks on specific IPs, making them
inaccessible~\cite{greatfire2015, china_Piratebay, fear_china}.

\begin{figure}[t]
    \centering
    \includegraphics[width=0.8\columnwidth]{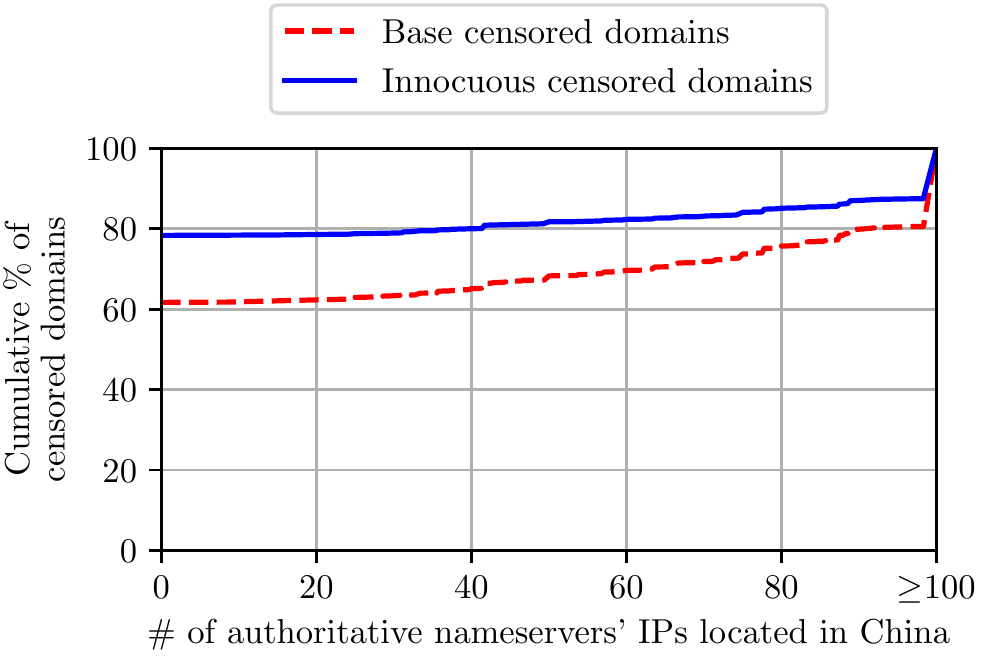}
    \caption{CDF of the number of authoritative name servers located inside
    China as a percentage of 138.7K base censored domains and 41K innocuously
    blocked domains.}
    \label{fig:cn_auth_ns}
    \vspace{-1.08em}
 \end{figure}

To estimate the extent to which the above geoblocking and overblocking
policies have impacted the global Internet, we analyze the location of
authoritative name servers of 138.7K base censored domains and 41K innocuously
blocked domains, using the MaxMind dataset~\cite{MaxMind}. As shown in
Figure~\ref{fig:cn_auth_ns}, 38\% (53K) of the base censored domains and
21.6\% (8.8K) of the innocuous censored domains have at least one
authoritative name server in China. In other words, there is always a non-zero
chance that DNS resolution for these 61.8K domains from outside China will be
poisoned, causing their visitors to potentially end up with an error page
similar to the above case. On the other hand, 19.4\% (26.9K) of base censored
domains and 12.5\% (5.1K) innocuously blocked domains have all of their
authoritative name servers in China, meaning that the resolutions for these
32K domains from outside China will always cross the GFW, thus being poisoned.

\subsection{Detection}
\label{sec:detection}

A common operational mechanism of DNS censorship is that the censor takes
advantage of the time-honored property of UDP-based DNS resolution to inject
poisoned responses, racing against the legitimate response. Depending on the
censored domain being queried, the GFW can even emit up to three responses.
This behavior of injecting multiple fake responses was first reported recently
in~\cite{Tripletcensors}. For the completeness of our investigation, we have
also identified the three different injectors based on the data collected by
\sysname, with more detailed analysis in
Appendix~\ref{sec:injector_fingerprint}.

From the GFW's perspective, the injection of multiple fake responses not only
increases the chance of successfully poisoning a censored client but also
makes it more costly and challenging to detect and circumvent its DNS
censorship~\cite{holdonDNS}. However, based on the pool of forged IPs and
their injection patterns that we have revealed in~\sectionref{sec:forged_ips},
detecting DNS censorship by the GFW can be done effectively by checking the
returned IP address against the pool of forged IPs discovered by \sysname.
Although this strategy may not detect all poisoned responses due to some rare
forged IPs that \sysname\ might have not observed in the long tail, from the
analysis of injection frequency in~\sectionref{sec:ip_frequency}, which we
have also verified its consistency across different network locations
(Appendix~\ref{sec:ip_consistency}), we are confident that this detection
technique can identify more than 99\% of the poisoned responses.

\begin{table}[t]
  \normalsize
  \centering
  \caption{Top ten public DNS resolvers with the highest number of censored
  domains whose poisoned resource records have polluted their cache.}
  \vspace{-0.8em}
  \resizebox{0.9\columnwidth}{!}{
  \begin{tabular}{rlrl}
  \toprule
  \textbf{\# Domains} &\textbf{Resolver} & \textbf{\# Domains} &
  \textbf{Resolver} \\
  \midrule
               74,715 &       Google     &              63,295 & OpenDNS \\
               71,560 &       Cloudflare &              62,825 & Comcast \\
               65,567 &       OpenNIC    &              56,913 & CleanBrowsing \\
               65,538 &       FreeDNS    &              56,628 & Level3 \\
               64,521 &       Yandex     &              55,795 & Verisign \\
  \bottomrule
  \end{tabular}}
  \label{tab:polluted_dns_resolvers}
  \vspace{-1.08em}
\end{table}

We next employ this detection technique to expose poisoned resource records
that have tainted public DNS resolvers around the world. In particular, once a
censored domain is detected by \sysname, we query them against popular DNS
resolvers and examine if its response matches any injection pattern we have
revealed in~\sectionref{sec:forged_ips}.
Table~\ref{tab:polluted_dns_resolvers} shows the top ten resolvers that have
been polluted with the highest number of censored domains. In total, we find
77K censored domains whose poisoned resource records have polluted the cache
of all popular public DNS resolvers that we examined. Of these
censored domains, 61K are base censored domains. This result aligns well with
our earlier speculation in~\sectionref{sec:impact}.

This finding shows the widespread impact of the bidirectional blocking
behavior of the GFW, necessitating the operators of these public DNS resolvers
to have an effective and efficient mechanism to prevent these poisoned
resource records from polluting their cache, to assure the quality of their DNS
service. Furthermore, the 61K base censored domains whose DNS queries from
outside China are censored is likely the reason why many censored domains
are classified as \emph{``newly observed domain''} or \emph{``not rated''}
in~\sectionref{sec:censored_domains_taxonomy}. This is because FortiGuard's
crawlers, which are likely located outside China, probably could not obtain
the correct IPs of these domains, thus failing to fetch and classify them.

%% file: circumvention.tex
\section{Circumvention}
\label{sec:circumvention}

We now show how insights gained from analyzing the censored domains
(\sectionref{sec:censored_domains}) and forged IPs discovered by \sysname\
over time (\sectionref{sec:forged_ips}) can assist us in developing strategies
to effectively and efficiently circumvent GFW's DNS censorship.

\subsection{Strategy}
\label{sec:strategy}

The GFW's bidirectional DNS filtering not only impacts in-China users but also
prevents users outside China from obtaining legitimate resources records of
geographically restricted domains based in China (\sectionref{sec:impact}).
Therefore, an effective DNS censorship evasion strategy would benefit not only
(1) users inside China who need to access censored domains hosted outside
China, but also (2) users outside China who need access to geoblocked domains
based in China. Both (1) and (2) also include open DNS resolvers located at
both sides of the GFW that want to prevent poisoned responses from polluting
their DNS cache.

Since the GFW operates as an on-path injector and does not alter the
legitimate response from the actual DNS resolver chosen by a client, a
circumvention strategy for the client is to not quickly accept any returned
responses when querying a censored domain. Instead, the client should wait for
an adjustable amount of time for all responses to arrive, as suggested
in~\cite{holdonDNS}. Upon receiving more than one IPv6 answer, the client can
filter out the bogus ones that belong to the Teredo subnet \texttt{2001::/32}.
Furthermore, for IPv4 answers, the client can check them against the injection
patterns and forged IPv4 addresses discovered in~\sectionref{sec:forged_ips}.

In our circumvention strategy, for each censored domain we need at least a
trustworthy resolver that possesses its genuine resource record(s). Popular
open resolvers (e.g., \texttt{8.8.8.8},  \texttt{1.1.1.1}) are often
considered as trustworthy sources when it comes to censorship evasion.
However, we have shown that the vast majority of public DNS resolvers have
been polluted with poisoned resource records (\sectionref{sec:detection}).
Therefore, we opt not to use them in this case, especially for obtaining the
legitimate resource records of geoblocked domains based in China. The only
remaining source that is immune to the GFW's poisoned responses and has a
given censored domain's genuine resource record(s) is its authoritative name
servers. This information is available in the zone files.

We send DNS queries for 138.7K base censored domains and 41K innocuous domains
to their authoritative name servers from our controlled machines located at both
sides of the GFW. We then expect to observe both censored and
non-censored resolutions at two sides of the GFW as a result of this
experiment. More specifically, from our US machine, resolutions for domains
whose authoritative name servers are located outside China will not be censored as
their queries will not cross the GFW, whereas resolutions for domains whose
authoritative name servers are located inside China are expected to be censored. On
the contrary, resolutions from our China machine towards authoritative
name servers located inside China will not be censored, while those queries sent
to authoritative name servers outside China will.

\subsection{Evaluation}
\label{sec:evaluation}

To evaluate the effectiveness of our method, we apply the proposed
circumvention strategy to filter out poisoned responses for those censored
resolutions and retain their ``legitimate'' responses, which we then compare
with actual legitimate responses returned from non-censored resolutions
conducted at the other side of the GFW. We find that our circumvention
strategy is highly effective, with an accuracy rate of 99.8\%. That is, 99.8\%
of responses classified as ``legitimate'' match the actual legitimate
responses obtained from non-censored resolutions. From a total of
1,007,002,451 resolutions that the GFW poisons, 1,005,444,476
responses classified as ``legitimate'' by our strategy contain the same
resource records (i.e., same IPs, CNAMEs, or IPs under the same AS for domains
hosted on Content Delivery Networks) with those observed from non-censored
resolutions. As discussed in~\sectionref{sec:ip_frequency}, there are a small
number of cases that we could not classify due to the invisibility of those
rarely injected forged IPs in the long tail that \sysname\ did not observe.
This finding highlights the importance of having an up-to-date and continuous
view into the pool of forged IPs for effectively circumventing the GFW's DNS
censorship.

To further assist in future adoptions of our strategy so that it will
not significantly downgrade the normal performance of other UDP-based DNS
resolutions for non-censored domains, we analyze the hold-on duration, which
the client should wait \emph{only} when resolving a censored domain, instead of
holding on for every resolution.

\begin{figure}[t]
   \centering
   \includegraphics[width=0.8\columnwidth]{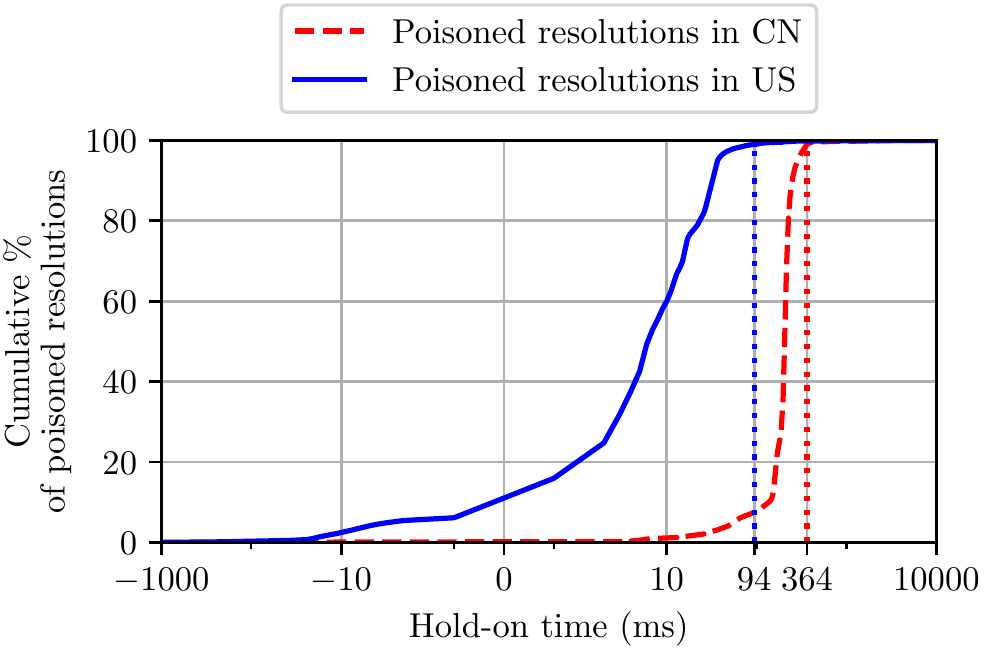}
   \caption{CDF of delta time between forged and legitimate responses measured
   from CN and US controlled machines.}
   \label{fig:holdon}
   \vspace{-1.08em}
\end{figure}

Figure~\ref{fig:holdon} shows the cumulative distribution of the delta time
between the first forged response and the legitimate one. The (red) dash line
is the CDF of the delta time measured at our China machine, and the (blue)
solid line is the CDF of this delta time measured at our US machine. On the
x-axis, a positive value means a poisoned response arrives before the
legitimate one. In contrast, a negative value indicates that the legitimate response
has arrived ahead of the fake ones.

As shown in the figure, the GFW can successfully poison more than 99.9\% of
all resolutions that carry censored domains, performed from our China machine
towards authoritative name servers located outside China. 99\% of poisoned
responses hit our machine within 364ms ahead of the legitimate ones. Although
this delta time may vary, depending on the relative distance between the
client and the GFW, for any client whose network location is close to ours,
this is the amount of extra time they should wait when resolving a censored
domain from inside China. In other words, upon receiving a DNS response after
querying a censored domain, the client should wait, at most, an extra 364ms
for the legitimate one to arrive. Users at different locations can
heuristically probe known censored domains to estimate the hold-on duration
that is representative for their location.

From the GFW's perspective, forged responses should ideally arrive at the
client before the legitimate one. From our US machine, we find that this is
not always true. Due to the unreliable and stateless nature of UDP packets
that might get lost or delayed when transferred between two distant locations,
and perhaps poisoning users outside China is not the primary design goal of
the GFW, 11\% of the poisoned responses arrive at our US machine after the
legitimate ones. Nevertheless, the remaining 89\% of fake responses still hit
our machine within 94ms ahead of the legitimate ones. This result again
highlights the importance of having a representative dataset of forged IPs
used by the GFW to effectively circumvent its DNS censorship. Especially when
fake responses arrive later, our dataset of forged IPs is necessary to avoid
misclassifying the legitimate ones arriving ahead as ``poisoned''.

\subsection{Analysis of True Resource Records}
\label{sec:trueRR}

Now that we have successfully obtained the legitimate resource records of the
138.7K base censored domains and 41K innocuously blocked domains, we next
analyze them to better understand the impact of blocking these domains. As
shown in Table~\ref{tab:trueRR}, 120K (86.8\%) base censored
domains have either an IPv4, IPv6, or CNAME resource record. In other
words, the remaining 18.7K (13.2\%) of the base censored domains that
currently do not have any resource records, indicating their inactivity. This
is also one of the reasons why we observe a large number of domains classified
as \emph{``newly observed domain''} and \emph{``not rated''} categories
in~\sectionref{sec:censored_domains_taxonomy}.

For the innocuously blocked domains, the actual impact of GFW's overblocking
may not be as severe because only 25.6K (62.5\%) of them have at least one
resource record. While the presence of resource records can be a sign of
(in)activeness for a given domain, it does not guarantee that a domain is
actively hosting any contents or services since a resource record can also be
used for redirecting visitors to a domain-parking site. Therefore, the total
number of domains with resource records shown in Table~\ref{tab:trueRR} should
be viewed as an upper bound of the actual number of domains that are actively
hosting any content or service. As part of our future work, we plan to visit
all of these domains using their true resource records and further investigate
the contents hosted on them.

Another focal point of Table~\ref{tab:trueRR} is the significantly high number
of CNAME resource records of both base censored domains and innocuously
blocked domains that have at least one authoritative name server located in China,
compared to domains whose authoritative name servers are located outside China. As
far as we are aware, this is because of a common workaround that is widely
suggested and used by domain owners who want to serve their websites to users
at both sides of the GFW since these CNAMEs are not filtered by the GFW.

\begin{table}[]
    \centering
    \caption{Breakdown of true resource records of base censored domains and innocuously blocked domains.}
    \resizebox{\columnwidth}{!}{
    \begin{tabular}{c|r|r|r|r|}
    \cline{2-5}
                                                                                         & \multicolumn{2}{c|}{Base censored domains}                          & \multicolumn{2}{c|}{Innocuously blocked domains}                \\ \hline
    \multicolumn{1}{|c|}{\multirow{2}{*}{\shortstack[l]{\# of domains by\\NS location}}} & \multicolumn{1}{c|}{$\geq$1 CN NS} & \multicolumn{1}{c|}{Non-CN NS} & \multicolumn{1}{c|}{$\geq$1 CN NS} & \multicolumn{1}{c|}{Non-CN NS} \\ \cline{2-5}
    \multicolumn{1}{|c|}{}                                                               & 53.1K (38.3\%)                     & 85.6K (61.7\%)                 & 8.9K (21.6\%)                      & 32.1K (78.4\%)                 \\ \hline
    \multicolumn{1}{|c|}{IPv4}                                                           & 29K (21.1\%)                       & 69.5K (50\%)                   & 6K (14.7\%)                        & 17.8K (43.5\%)                 \\ \hline
    \multicolumn{1}{|c|}{IPv6}                                                           & 1.3K (1\%)                         & 28K (20.2\%)                   & 0.1K (0.3\%)                       & 2.8K (7\%)                     \\ \hline
    \multicolumn{1}{|c|}{CNAMEs}                                                         & \textbf{31K (22.3\%)}              & 3.6K (2.6\%)                   & \textbf{2.9K (7.1\%)}              & 0.5K (1.3\%)                   \\ \hline
    \multicolumn{1}{|c|}{\shortstack[l]{\# of domains\\with RR(s)}}                      & \multicolumn{2}{c|}{120K (86.8\%)}                                  & \multicolumn{2}{c|}{25.6K (62.5\%)}                            \\ \hline
\end{tabular}}
    \label{tab:trueRR}
    \vspace{-1.08em}
    \end{table}

%% file: discussion.tex
\section{Discussion}
\label{sec:discussion}

In this section, we discuss the limitations of our study and provide
suggestions for involving parties that are impacted by the GFW's DNS
censorship.

\subsection{Limitations}
\label{sec:limitations}

In order to compare our analysis on the categories of censored domains with
prior studies, we choose to use a common classification service provided by
FortiGuard~\cite{FortiGuard}. However, we discovered that the GFW's
overblocking and geoblocking policy could have already impacted this service
(\sectionref{sec:detection}). Moreover, Vallina et
al.~\cite{Vallina2020MisshapesMM} have shown that different classification
services could result in different views of the domains being categorized. We
thus tried to obtain additional classification services from two other
vendors, namely, McAfee and VirusTotal. However, we were told by
McAfee~\cite{McAfee} that they only provide the service for business
customers, and VirusTotal~\cite{VirusTotal} did not respond to our requests.

Similar to other studies in remote censorship
measurement~\cite{VanderSloot2018QuackSR, Raman2020MeasuringTD,
Raman2020CensoredPlanet}, packets sent from our measurement infrastructure may
get blocked or discriminated by the GFW. However, over the course of more than
nine months operating \sysname, we did not experience any disruptions caused
by such discriminative behaviors, as is evident by the consistency observed
between the data collected by \sysname\ and across different network locations
(Appendix~\ref{sec:ip_consistency}). Moreover, as part of our outreach
activities, we have also received confirmations from local Chinese advocacy
groups and owners of censored domains detected by GFWatch when reaching out to
these entities to share our findings. Nonetheless, if our measurement machines
ever gets blocked, we can always dynamically change their network location.

Finally, we develop \sysname\ as a measurement system to expose the GFW's
blocking behavior based on DNS censorship. However, this is not the only
filtering technique used by the GFW; censorship can also happen at other
layers of the network stack, as previously studied~\cite{Barme1997,
Zittrain2003, Clayton2006IgnoringTG, Park2010EmpiricalSO, Winter2012,
ensafi2015analyzing, Arun:foci18}. Although prior works have shown that some
websites could be unblocked if the actual IP(s) of censored domains can be
obtained properly~\cite{hoang:2019:measuringI2P, Chai2019OnTI}, securing DNS
resolutions alone may not be enough in some cases because blocking can also
happen at the application layer (e.g., SNI-based blocking~\cite{Chai2019OnTI},
keyword-based filtering~\cite{Rambert2021ChineseWO}) or even at the IP
layer~\cite{Hoang2018:IMC, Hoang2021:PoPETS}, regardless of potential
collateral damage~\cite{Hoang2020:CCR}.

Nonetheless, DNS is one of the most critical protocols on the Internet since
almost every online communication starts with a DNS lookup. We believe that
continuously monitoring the GFW's filtering policy at this layer is necessary
and important to timely inform the public of the erratic changes in China's
information controls policies, both from technical and political perspectives.
Appendix~\ref{sec:political_censorship} provides some examples of domains
censored due to political motivations.

\subsection{Suggestions}
\label{sec:suggestions}

\myparab{GFW operators.} Although the widespread impact of the GFW's DNS
filtering policy is clear, as shown throughout this paper,
we are not entirely certain whether this censorship policy is intentional or
accidental. While prior works have shown intermittent failures of the
GFW~\cite{ensafi2015analyzing, Tripletcensors}, all geoblocking of China-based
domains and overblocking of innocuous domains discovered by \sysname\ have
lasted over several months. This relatively long enough period of time leads
us to believe that the GFW's operators would have clearly known about the
global impact of their DNS filtering policy. By exposing these negative
impacts on several parties outside China to the public, we hope to send a
meaningful message to the GFW's operators so that they can revise their DNS
filtering policy to reduce its negative impacts beyond China's borders.

\myparab{Public DNS resolvers.} Poisoned DNS responses have widely polluted
all popular public DNS resolvers outside China due to the geoblocking and
overblocking of many domains based in China
(\sectionref{sec:censorship_leakage}). DNSSEC~\cite{rfc2065} has been
introduced to assure the integrity and authenticity of DNS responses for more
than two decades to address these problems. However, DNSSEC is not widely
adopted because of compatibility problems and technical
complications~\cite{Dai:CANS16:DNSSEC, Chung:Usenix17:DNSSEC,
Hao:Usenix18:CDNSEC}. To this end, public DNS resolvers can use the strategy
introduced in~\sectionref{sec:circumvention} to prevent poisoned DNS responses
spoofed by the GFW from tainting their cache. By waiting for all responses to
arrive and comparing the answers with the pool of forged IPs discovered by
\sysname\ (\sectionref{sec:forged_ips}), public DNS resolvers can filter out
99\% of poisoned responses by the GFW. Note that it is not always necessary to
wait for all responses to arrive because the GFW does not censor all domains.
As we will make both censored domains and forged IPs publicly available and
update them on a daily basis, these datasets can be used to decide whether to
wait or not when resolving a given domain. This way, public DNS resolvers
would be able to prevent poisoned responses from polluting their cache,
assuring the quality of their DNS service while avoiding any downgrades of
normal performance when resolving domains that are not censored.

\myparab{Owners of forged IPs.} Legitimate owners of forged IPs may try to
avoid hosting critical services on these IPs as their resources may be
saturated due to handling unsolicited TCP and HTTP(s) requests, as shown
in~\sectionref{sec:impact}. Currently, we do not find evidence that the GFW is
using these forged IPs as a way to saturate computing resources of the
infrastructure behind them since there are more than 1.7K forged IPs in the
pool (\sectionref{sec:forged_ips_over_time}) and most of them are dynamically
injected (\sectionref{sec:ip_frequency}). However, a previous report of the
Great Cannon~\cite{Marczak2015AnAO} has shown that China is willing to
weaponize the global Internet to mount resource exhaustion attacks on specific
targets. With DNS censorship, the GFW can adjust its injection pattern to
concentrate on a handful of forged IPs, resulting in a large amount of
requests towards these targeted IPs and thus saturating their computing
resources~\cite{greatfire2015, china_Piratebay, fear_china}.

\myparab{Domain owners.} Using our dataset of censored domains, domain owners
can check whether their domain is censored or not, and censored due to
intended blocking or overblocking. Unless the GFW's operators revise their
blocking rules, future domain owners should try to refrain from registering
domains that end with any overblocking patterns discovered
in~\sectionref{sec:reverse_engineer_blocklist} to avoid them being
inadvertently blocked by the GFW.

\myparab{End users.} Despite the large number of censored domains discovered
by \sysname, different Internet users may be interested in different subsets
of these censored domains, but not all. As an immediate countermeasure to the
GFW's DNS censorship, we will make the legitimate resource records of censored
domains obtained in~\sectionref{sec:circumvention} publicly available on a
daily basis. This way, impacted users can look up and store legitimate
resource records for particular censored domains in their system's
\texttt{hosts} file to bypass the GFW's DNS censorship. Alternatively, a
censorship-circumvention component of software can implement the hold-on
strategy (\sectionref{sec:circumvention}) and gather records based on the
client's location. In case the client cannot access the sanitized data
published by \sysname, another client-side strategy is to send two
back-to-back queries. Depending on whether a censored domain belongs to the
dynamic or static injection groups (\sectionref{sec:ip_frequency}), the client
can discern which responses are legitimate. Since the majority of censored
domains are poisoned with dynamic IPs, the client can classify the legitimate
responses, which typically point to the same IP (due to back-to-back queries)
or the same AS. This way, the software only needs to know whether its intended
domains are poisoned with static or dynamic IPs. To this end, continuous
access to GFWatch's data is not necessary for this strategy to work, while
fresh records can still be obtained.

%% file: related.tex
\section{Related Work}
\label{sec:related}

In addition to~\cite{Tripletcensors}, which is the most recent work related to
ours that we have provided in-depth discussions throughout our paper, some
other one-time studies have also looked into the DNS censorship behavior of
the GFW in the past~\cite{Zittrain2003, Tokachu, Lowe2007a, brown2010dns,
Sparks:CCR2012, china:2014:dns:anonymous}. While China's GFW may not be the
primary and sole focus, there are platforms actively measuring censorship
around the globe that may also have a partial view into the GFW's DNS
censorship behavior~\cite{ICLab:SP20, filasto2012ooni,
Raman2020CensoredPlanet}. To provide our readers with a complete view of these
efforts and highlight how our study is different from them, we summarize the
major differences among these studies in this section. A more detailed
comparison table can be found in Appendix~\ref{sec:related_table}.

In its early days, the GFW only used a handful of forged
IPs~\cite{Zittrain2003, Lowe2007a}. However, later studies have noticed an
increase in the number of forged IPs, from nine in 2010~\cite{brown2010dns},
28 in 2011~\cite{Sparks:CCR2012}, 174 in 2014~\cite{china:2014:dns:anonymous},
to more than 1.5K recently~\cite{Tripletcensors}. Except
for~\cite{Sparks:CCR2012} and~\cite{china:2014:dns:anonymous} whose authors
preferred to remain anonymous and the dataset URLs provided in their papers
are no longer accessible, we were able to obtain data from other studies for
comparison (Table~\ref{tab:comparison_table}). A common drawback of these
studies is that their experiments are conducted only over limited time periods
and the test domains are also static, i.e., obtained from a snapshot of Alexa
top list or zone files.

To address this drawback of previous one-off studies, longitudinal platforms
have been created to measure censorship around the world, including
ICLab~\cite{ICLab:SP20}, OONI~\cite{filasto2012ooni}, and Censored
Planet~\cite{Raman2020CensoredPlanet}. To reduce risks to volunteers and
observe interferences at multiple layers of the network stack,
ICLab~\cite{ICLab:SP20} chooses commercial VPNs as vantage points for their
measurement. However, this design choice limits their visibility into China as
commercial VPNs are restricted in the country~\cite{Bloomberg2017,
Reuters2017}. With different approaches, OONI~\cite{filasto2012ooni} recruits
volunteers to participate in censorship measurements, whereas Censored
Planet~\cite{Raman2020CensoredPlanet} employs a series of remote measurement
techniques to infer censorship. These design choices allow the two later
platforms to obtain vantage points located in China for their measurements. We
fetch data collected during the same period of our study available on these
projects' websites for comparison.

For OONI data, we first gather measurements conducted by volunteers in China
that are flagged as ``DNS inconsistency''~\cite{oonidns}. To reduce false
positives due to domains hosted on CDNs, we filter out those cases where
controlled and probed responses have different IPs but belong to the same AS.
After sanitization, we find 710 forged IPs from OONI data, 593 of which are in
common with those observed by \sysname. Examining the different cases, we find
that there are still misclassified cases due to domains hosted on popular CDNs
whose network spans across different AS numbers.

For Censored Planet~\cite{Raman2020CensoredPlanet}, we use data collected by
the Satellite~\cite{Satellite} module for comparison since it is designed to
measure DNS-based network interference. Satellite infers DNS censorship by
comparing responses received from open DNS resolvers with ones obtained from a
control resolver, along with other metadata such as AS number, HTTP static
content, and TLS certificates. Since Satellite's data is not annotated with
geographical information, we use different geolocation datasets~\cite{MaxMind,
ip2location, ipinfo, dbip} to confirm the location of open resolvers used by
Satellite. We then extract responses from open resolvers located in China that
are flagged as \texttt{``anomaly''}. We find a total of 2.4K forged IPs
reported by Satellite, 1.6K of which are in common with ours. The difference
in the number of forged IPs in this case, is due to the inherent nature of
Satellite's measurement approach of using open DNS resolvers. In particular,
about 600 IPs observed by Satellite, but not \sysname, belong to Cisco
OpenDNS, which provides DNS-based network filtering services for various
customer types, ranging from home to business users~\cite{cisco_openDNS}. From
a detection point of view, these censorship cases are valid, but due to
different local policies of these open resolvers, instead of country-level
censorship enforced by the GFW.

A shared property of OONI and Satellite is that measurement vantage points
(volunteers' devices and open resolvers) are not owned by these platforms.
Therefore, only a limited number of domains can be tested with adequate
frequency to avoid saturating these vantage points' computing resources. To
overcome this pitfall, \sysname's measurement approach of using our own
machines located at both sides of the GFW allows us to test hundreds of
millions of domains multiple times per day. Using machines under our control
also reduces the false positive rate to \emph{zero} since neither of our
machines have any DNS resolution capabilities.

%% file: conclusion.tex
\section{Conclusion}
\label{sec:conclusion}

In this work, we develop \sysname, a large-scale longitudinal measurement
platform, to provide a constantly updated view of the GFW's DNS-based blocking
behavior and its impact on the global Internet. Over a nine-month period,
\sysname\ has tested 534M domains and discovered 311K censored domains.

We find that the GFW's DNS censorship has a widespread negative impact on the
global Internet, especially the domain name ecosystem. \sysname\ has detected
more than 77K censored domains whose poisoned resource records have polluted
many popular public DNS resolvers, including Google and Cloudflare. Based on
insights gained from the data collected by \sysname, we then propose
strategies to effectively detect poisoned responses and evade the GFW's DNS
censorship.

As \sysname\ continues to operate, our data will not only cast new light on
technical observations, but also timely inform the public about changes in the
GFW's blocking policy and assist other detection and circumvention efforts.

%% file: ack.tex
\section*{Acknowledgments}

We are grateful to Ronald J. Deibert, Adam Senft, Lotus Ruan, Irene Poetranto,
Hyungjoon Koo, Shachee Mishra, Tapti Palit, Seyedhamed Ghavamnia, Jarin Firose
Moon, Md Mehedi Hasan, Thai Le, Eric Wustrow, Martin A. Brown, Siddharth
Varadarajan, Ananth Krishnan, Peter Guest, and others who preferred to remain
anonymous for helpful discussions and suggestions.

We would like to thank all the anonymous reviewers for their thorough feedback
on this paper. We especially thank the team at \texttt{GreatFire.org} for
helping to share our findings with related entities in a timely fashion.

This research was supported by the Open Technology Fund under an Information
Controls Fellowship. The opinions in this paper are those of the authors and
do not necessarily reflect the opinions of the sponsor.

%% file: appendix.tex
\appendix

\section{Most Extreme Blocking Rules}
\label{sec:extreme_blocking rules}

\begin{table}[t]
      \centering
      \caption{Top base censored domains that cause most overblocking of innocuous domains.}
      \resizebox{\columnwidth}{!}{
      \begin{tabular}{rlr}
      \hline
      \small{\# domains}          & \small{Base censored domains}   & \small{Sample innocuous domains} \\
      \small{impacted}            &                                 & \\
      \hline
     11,227 & \texttt{919.com}    & \texttt{455919.com}, \texttt{rem99919.com}      \\
            &                     & \texttt{niwa919.com}, \texttt{xaa919.com}     \\
      \hline
      2,346 & \texttt{jetos.com}  & \texttt{ccmprojetos.com}, \texttt{csprojetos.com}\\
            &                     & \texttt{itemsobjetos.com}, \texttt{dobobjetos.com}     \\
      \hline
      1,837 & \texttt{33a.com}    & \texttt{87833a.com}, \texttt{280333a.com}      \\
            &                     & \texttt{xn---72caa7c0a9clrce0a1fp33a.com}     \\
            &                     & \texttt{xn---zck4aye2c2741a5qvo33a.com}      \\
      \hline
      1,574 &\texttt{9444.com}    & \texttt{mkt9444.com}, \texttt{15669444.com}    \\
            &                     & \texttt{3329444.com}, \texttt{5719444.com}     \\
      \hline
      1,547 &\texttt{sscenter.net}& \texttt{dentalwellnesscenter.net}, \texttt{swisscenter.net}      \\
            &                     & \texttt{chesscenter.net}, \texttt{childlosscenter.net}      \\
      \hline
      1,487 & \texttt{1900.com}   & \texttt{faber1900.com}, \texttt{salah1900.com}      \\
            &                     & \texttt{phoenixspirit1900.com}, \texttt{interiors1900.com}    \\
      \hline
      1,392 & \texttt{98a.com}    & \texttt{p98a.com}, \texttt{72898a.com}, \texttt{1098a.com}       \\
            &                     & \texttt{xn---1-ieup4b2ab8q5c0dxj6398a.com}     \\
      \hline
      1,144 & \texttt{ss.center}  & \texttt{hss.center}, \texttt{icass.center}      \\
            &                     & \texttt{limitless.center}, \texttt{ass.center}     \\
      \hline
      1,089 & \texttt{reddit.com} & \texttt{bestiptvreddit.com}, \texttt{booksreddit.com}      \\
            &                     & \texttt{cachedreddit.com}, \texttt{geareddit.com}     \\
      \hline
        789 & \texttt{visi.tk}    & \texttt{erervisi.tk}, \texttt{yetkiliservisi.tk}      \\
            &                     & \texttt{buderuservisi.tk}, \texttt{bodrumklimaservisi.tk}     \\
        \bottomrule
  \end{tabular}}
  \vspace{-1em}
  \label{table:innocuous_censored_domains}
  \end{table}

Table~\ref{table:innocuous_censored_domains} shows the top ten base censored
domains blocked under Rule 4 that we have discussed
in~\sectionref{sec:reverse_engineer_blocklist}. The blocking rule applied on
these ten domains results in overblocking of more than 24K innocuous domains,
which is more than half of all innocuous domains. The third column shows some
samples of innocuous censored domains that \sysname\ has discovered. The
impacted innocuous domains presented in this table are all active and hosting
some contents at the time of writing this paper. Except those that do not
allow Web Archive's crawler, we have also saved a snapshot of these domains at
\texttt{https://web.archive.org} for future reference in case these domains
become inactive. In contrast, most base censored domains shown in the second
column are not currently hosting any content. Therefore, one may wonder why
many seemingly inconsequential domains are being censored.

To make sure that these seemingly inconsequential censored domains were not
blocked because the GFW was using an imprecise classifier (e.g., a Bloom
filter) for fast classification, we tested 200M randomly generated nonexistent
domains and found that none were censored. It is worth noting that many
censored domains discovered by \sysname\ have been blocked before the launch
of our platform. Prior to our testing, they might have served ``unwanted''
content that we were not aware of. Moreover, the GFW is known to conduct
blanket blocking against websites that run editorials on ``unwanted'' topics
without carefully verifying their contents. Once domains are censored, they
are often kept in the GFW's blocklist for a long time regardless of their
activity~\cite{China_jawapos}.

As can be seen from the table, the GFW's overblocking design affects not only
usual ASCII-based innocuous domains, but also Internationalized Domain Names
(IDNs), i.e., those starting with \texttt{``xn---''}. Of 41K innocuously
blocked domains, we find a total of 1.2K IDNs are overblocked. Our finding
shows that the current DNS-based blocking policy of the GFW has a widespread
negative impact on the domain name ecosystem.

\section{Consistency of Forged IP Addresses Across Different Network Locations}
\label{sec:ip_consistency}
\begin{figure}[t]
      \centering
      \includegraphics[width=.9\columnwidth]{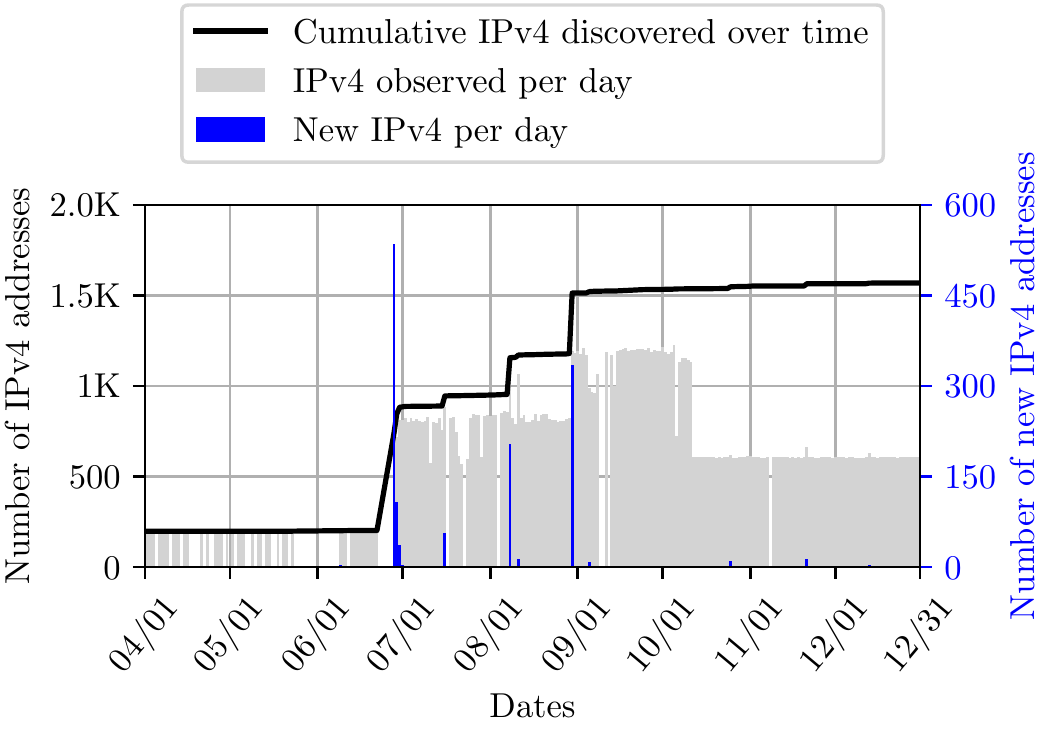}
      \caption{Number of forged IPv4 addresses detected over time by probing
      different network prefixes in China.}
      \label{fig:prefix_probing_ipv4_overtime}
\end{figure}

To confirm whether the pool of forged IPs discovered by \sysname\
(\sectionref{sec:forged_ips}) is representative enough, we probe different
network locations in China to compare the forged IPs observed from these
locations and the ones seen by \sysname. For this experiment, we obtain the
daily updated \emph{pfx2as} dataset provided by CAIDA~\cite{CAIDA:pfx2as}, and
extract prefixes located in China by checking them against the MaxMind
dataset~\cite{MaxMind}, which we also update biweekly. Unlike the measurement
conducted between our own controlled machines located at two sides of the GFW,
this task requires us to send DNS queries, encapsulating censored domains, to
destinations we do not own. Although similar large-scale network probing
activities are widely conducted nowadays by both
academia~\cite{Durumeric2013ZMapFI, Iris, VanderSloot2018QuackSR,
Raman2020CensoredPlanet} and industry~\cite{shodan, rapid7}, our measurement
must be designed in a careful and responsible manner.

Our sole purpose of this measurement is to deliver probing queries passing
through the GFW's infrastructure at different network locations to trigger
censorship, instead of having the probing packets completely delivered to any
alive hosts. Therefore, we craft our probing packets using the routing address
of a given prefix as the destination IP. According to the best current
practice~\cite{rfc4632}, except for the case of a \texttt{/32} subnet with
only one IP, the routing address of a subnet should not be assigned to any
device because it is solely used for routing purposes. For example, given the
prefix \texttt{1.92.0.0/20} announced in the \emph{pfx2as} dataset, we craft
our probing packet with the destination as \texttt{1.92.0.0}. With this
probing strategy, we can reduce the risk that our packets will hit an alive
host while still being able to deliver them across the GFW's infrastructure at
different network locations. To reduce the risk even further, we opt to only
probe prefixes whose subnet is less-specific than \texttt{/24}.

In spite of the standardized practices in assigning IP and the extra care that
we have taken in designing our measurement, we also follow a common practice
that is widely used in research activities that involve network scanning,
i.e., allowing opt-out. More specifically, we accompany our probing DNS
queries with a non-censored domain under our control, from which the
information about our study and a contact email address can be found to
request opt-out from our measurement. Since the launch of \sysname, we have
not received any complaints or opt-out requests.

Figure~\ref{fig:prefix_probing_ipv4_overtime} show the cumulative number of
forged IPs discovered daily and over the whole period of our measurement.
Similar to Figure~\ref{fig:ipv4_overtime}, the number of forged IPs addresses
observed initially in April is also about 200. However, we did not see any
gradual increase in the number of forged IPs from May as seen in
Figure~\ref{fig:ipv4_overtime}. After waiting about two months without seeing
any new IPs observed from probing different prefixes, we have learned that
this is due to the fact that we only use \emph{one} known censored domain for
probing the prefixes. This is because of an earlier precaution that these
probed destinations are not owned by us, thus we should try to limit the
amount of probing traffic as much as possible. However, it turned out that we
need to probe more than just one domain to be able to obtain a similar set of
forged IP addresses detected earlier by \sysname.

We then decide to add more domains to this test, probing a total of 22
censored domains per prefix. These domains are selected from several
categories, including advocacy organizations, proxy avoidance, news and media,
social network, personal websites and blogs, shopping, instant messaging, etc.
As expected, the cumulative number of forged IPs immediately increases to
almost 1K the day we revise our test domains. Similar to
Figure~\ref{fig:ipv4_overtime}, the cumulative number of forger IPs also
increase gradually towards the end of August. With a major increase of more
than 300 forged IPs, the number of all forged IPs observed from our prefixes
probing measurement also converges to above 1.5K by the end of December.

While the number of forged IPs obtained from probing the prefixes on some
days, especially from July to September, is higher than what \sysname\
observed during this period, we find that 96\% of the forged IPs observed from
prefixes probing have already detected by \sysname. Conducting the same
injection frequency analysis on these forged IPs gives us the same results as
found in~\sectionref{sec:ip_frequency}. In other words, the most frequently
injected IPs discovered by \sysname\ and from probing different prefixes are
the same. To this end, we could confirm that the coverage of forged IPs
discovered by \sysname\ is representative and sufficient for us to develop
effective detection (\sectionref{sec:detection}) and circumvention strategies
(\sectionref{sec:circumvention}).

\section{Multiple Injectors}
\label{sec:injector_fingerprint}

\begin{table*}[t]
      \normalsize
      \centering
      \caption{A high-level comparison of censored domains and forged IPs
      detected by different studies/platforms. (*) The number of forged IPs
      from Satellite and OONI includes ``anomalies'' due to domains hosted on
      CDNs and localized filtering policies.}
      \resizebox{1.7\columnwidth}{!}{
      \begin{tabular}{llcrrrr}
      \toprule
      \textbf{Study/Platform}                 &\textbf{Duration}    & \textbf{Longitudinal} & \textbf{Tested}  & \textbf{Censored} & \textbf{Forged} & \textbf{Common} \\ [0.5ex]
                                              &                     &                       & \textbf{Domains} & \textbf{Domains}  & \textbf{IPs}    & \textbf{Forged IPs} \\ [0.5ex]
      \midrule
      Zittrain et al.~\cite{Zittrain2003}     & Mar 2002 - Nov 2002 &  $\Circle$            & 204K             & 1K                & 1               & 1 \\ 
      Lowe et al.~\cite{Lowe2007a}            & 2007                &  $\Circle$            & 951              & 393               & 21              & 3 \\
      Brown et al.~\cite{brown2010dns}        & Nov 2010            &  $\Circle$            & 1                & 1                 & 9               & 6 \\
      CCR'12~\cite{Sparks:CCR2012}            & Nov 2011            &  $\Circle$            & 10               & 6                 & 28              & \\
      FOCI'14~\cite{china:2014:dns:anonymous} & Aug 2013 - Apr 2014 &  $\Circle$            & 130M             & 35.3K             & 174             & \\
      Triplet Censors~\cite{Tripletcensors}   & Sep 2019 - May 2020 &  $\Circle$            & 1M               & 24.6K             & 1,510           & 1,462\\
      OONI~\cite{filasto2012ooni}             & Apr 2020 - Dec 2020 &  $\CIRCLE$            & 3.3K             & 460               & \mbox{*}710     & 593 \\
      Satellite~\cite{Satellite}              & Apr 2020 - Dec 2020 &  $\CIRCLE$            & 3.5K             & 375               & \mbox{*}2,391   & 1,613 \\
      \sysname                                & Apr 2020 - Dec 1020 &  $\CIRCLE$            & 534M             & 311K              & 1,781           &  - \\
      \bottomrule
      \end{tabular}}
      \label{tab:comparison_table}
    \end{table*}

It was first reported by~\cite{Tripletcensors} that the GFW comprises multiple
injectors that are responsible for DNS poisoning. Depending on the domain
being queried (e.g., \texttt{google.sm}), multiple forged responses can be
triggered simultaneously to increase the chance of successfully poisoning
censored clients if one of the injectors is overloaded, and make detection and
circumvention non-trivial. From the data collected by \sysname, we have
confirmed the same injection behavior. More specifically, there are three
injectors, which can be differentiated by the \emph{``DNS Authoritative
Answer''} flag in the DNS header and the \emph{``do not fragment''} flag in
the IP header. Injector 1 has the \emph{``DNS Authoritative Answer''} bit set
to \textbf{1}, Injector 2 has the \emph{``DNS Authoritative Answer''} bit set
to \textbf{0} and \emph{``do not fragment''} bit set to \textbf{1}, whereas
Injector 3 has the \emph{``DNS Authoritative Answer''} bit set to \textbf{0}
and \emph{``do not fragment''} bit set to \textbf{0}.

\begin{figure}[t]
      \centering
      \includegraphics[width=0.9\columnwidth]{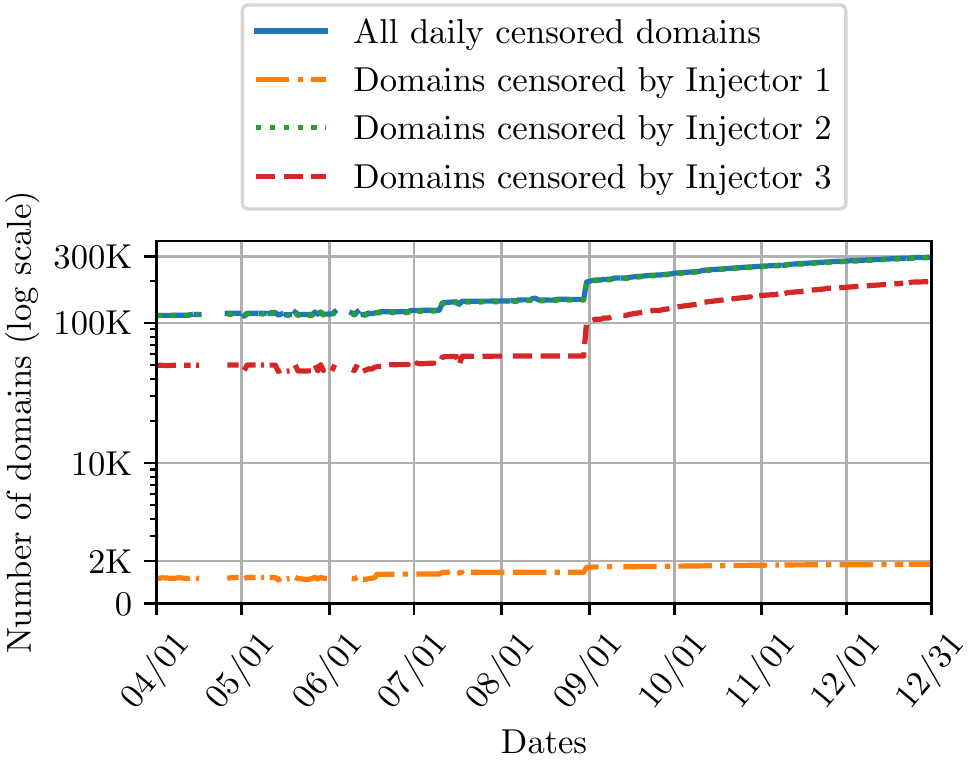}
      \caption{Number of censored domains per injector.}
      \label{fig:injector_domain}

\end{figure}

Based on these fingerprints, we then cluster 311K censored domains into three
groups with respect to the three injectors. Figure~\ref{fig:injector_domain}
depicts the number of censored domains observed over time for each injector.
Injector 2 is responsible for 99\% of the censored domains, whereas Injectors
3 and 1 are responsible for only 64\% and less than 1\% (2K) of censored
domains, respectively. Note that all domains censored by Injector 3 are also
censored by Injector 2, while there are 1.7K domains censored only by Injector
1, but not other injectors.

\section{Politically Motivated Censorship}
\label{sec:political_censorship}

Internet censorship and large-scale network outages are often politically
motivated~\cite{gill2015characterizing, Richter2018AdvancingTA}. From the
censored domains discovered by \sysname, we find numerous governmental
websites censored by the GFW, including many sites belonging to the US
government, such as \texttt{share.america.gov}, \texttt{cecc.gov}, and
\texttt{uscirf.gov}.

During the nine-month measurement period, \sysname\ has also spotted several
blockages that coincide with political events. For instance, soon after the
clash between China and India due to the border dispute in
Ladakh~\cite{IN-CN-clash}, on June 18th 2020 \sysname\ detected the DNS
filtering of several Indian news sites (e.g., \texttt{thewire.in},
\texttt{newsr.in}). We reached out to the editor of the Wire India to report
blockage against their website by the GFW and were told that they were unaware
of the blockage since the site was still accessible from China earlier.
Another instance is the blockage of \texttt{scratch.mit.edu} that took place
in August, 2020, due to some content deemed as anti-China hosted on this
website, affecting about three million Chinese users~\cite{mit_scratch_ban}.
Although this event was reported by the GreatFire
project~\cite{greatfire_project} on the 20th and by Chinese users on the
14th~\cite{mit_scratch_ban}, \sysname\ actually detected the first DNS
poisoning instance earlier on August 13th.

These cases highlight the importance of \sysname's ability to operate in an
automated and continuous fashion to obtain a constantly updated view of the
GFW to timely inform the public about changes in its blocking policy.

\section{Detailed Comparison with Related Work}
\label{sec:related_table}

Table~\ref{tab:comparison_table} provides a detailed comparison, highlighting
the main differences between \sysname\ and prior studies. Note that the
numbers of IPs in this table indicate IPv4 addresses. We do not include a
comparison of the number of IPv6 addresses because most previous works did not
consider IPv6 in their experiments.